\documentclass[%
twocolumn,
superscriptaddress,
 amsmath,amssymb,aps, 
 floatfix, showpacs,
prb,
citeautoscript,
longbibliography,
]{revtex4-1}

\usepackage{graphicx}
\usepackage{dcolumn}
\usepackage{bm}
\usepackage{amsfonts,amssymb,amsmath}
\usepackage{hyperref}
\usepackage{braket}
\usepackage{color}
\usepackage{soul}
\begin{document}

\preprint{}

\title{Lateral Interfaces of Transition Metal Dichalcogenides: A Stable Tunable One-Dimensional Physics Platform}

\author{Oscar \'Avalos-Ovando}
 \email{oa237913@ohio.edu}
 \affiliation{Department of Physics and Astronomy, and Nanoscale and Quantum Phenomena Institute, \\ Ohio University, Athens, Ohio 45701--2979, USA}
\author{Diego Mastrogiuseppe}
\affiliation{Instituto de F\'isica Rosario (CONICET), 2000 Rosario, Argentina}
\author{Sergio E. Ulloa}
\affiliation{Department of Physics and Astronomy, and Nanoscale and Quantum Phenomena Institute, \\ Ohio University, Athens, Ohio 45701--2979, USA}

\date{\today}

\begin{abstract}
We study in-plane lateral heterostructures of commensurate transition-metal dichalcogenides, such as MoS$_{2}$-WS$_{2}$ and MoSe$_{2}$-WSe$_{2}$, and find interfacial and edge states that are highly localized to these regions of the heterostructure.
These are one-dimensional (1D) in nature, lying within the bandgap of the bulk structure and exhibiting complex orbital and spin structure.
We describe such heteroribbons with a three-orbital tight-binding model that uses first principles and experimental parameters as input, allowing us to model realistic systems.
Analytical modeling for the 1D interfacial bands results in long-range hoppings due to the hybridization along the interface, with strong spin-orbit couplings.
We further explore the Ruderman-Kittel-Kasuya-Yosida indirect interaction between magnetic impurities located at the interface.  The unusual features of the interface states result in effective long-range exchange non-collinear interactions between impurities.
These results suggest that transition-metal dichalcogenide interfaces could serve as stable, tunable 1D platform with unique properties
for possible use in exploring Majorana fermions, plasma excitations and spintronics applications.
\end{abstract}

\maketitle

\section{Introduction}
\label{sec:introduction}

Research on two-dimensional (2D) transition metal dichalcogenides (TMDs)\cite{Manzeli20172d} has been growing rapidly since the isolation of semiconducting monolayers in 2005\cite{Novoselov2005}, and specially after the discovery of their direct band gap in the monolayer limit\cite{Mak2010}. A TMD monolayer results in a $MX_2$ trigonal prismatic environment ($M$=Mo, W, and $X$=S, Se, Te) where
two chalcogen layers sandwich a transition metal one. This structure and large intrinsic spin-orbit coupling (SOC)\cite{Zhu2011,Xiao2012,Xu2014NatPhys} gives rise to interesting spin-valley coupling and polarization-dependent optical response\cite{Xiao2012,Xu2014NatPhys}. Moreover, progress in the synthesis of TMDs has allowed the combination of different low-dimensional materials, creating interesting heterostructures (HSs). These HSs have received a lot of attention lately since they are capable of enhancing or, better yet, creating new tailored features, which are rather weak or nonexistent in their pristine counterparts. Prominent recent examples include enhancement of valley splitting by magnetic proximity effects\cite{Zhao2017enhanced,Seyler2018}, the appearance of spatially indirect excitons\cite{Calman2018}, and superconductivity in graphene superlattices\cite{Cao2018unconventional}. While most of current research is based on stacked (or van der Waals) HSs\cite{Geim2013,Novoselov2016}, attention has been also focused on {\em lateral} HSs, with two different 2D materials joined to form a 1D interface. Examples of those systems include graphene-hexagonal boron nitride (hBN)\cite{Drost2015graphenehBN}, graphene-TMDs\cite{Xi2016parallel}, hBN-TMDs\cite{Xi2016parallel}, and different TMD-TMD combinations\cite{Huang2014,Gong2014,Duan2014,Zhang2015,Li2015,Zhang2016naturecommunications,Xi2016parallel,Zhang2018strain,Sahoo2018,Xie2018ParkGroup}, with many suggested applications as in-plane transistors, diodes, p-n photodiodes and CMOS inverters.

Experiments in this area have focused successfully on improving the lateral atomic connection between the two materials, in order to build a clean and sharp border between them. The progress is clearly reflected in the literature, as the description has changed from \emph{alloy} to \emph{interface} to describe the lateral junctures. The interface between both materials can be an exciting new platform for the study of 1D physics. In TMDs, chemical vapor deposition (CVD) growth techniques\cite{Huang2014,Gong2014,Duan2014} of lateral HSs have allowed very sharp and well oriented interfaces. The HSs achieved are usually triangular flakes composed of a central TMD material and an outer triangular ring of a different TMD. These are grown by changing conditions during the growth process, such as keeping the chalcogen $X$ fixed and changing the metal $M$ resulting in MoSe$_{2}$-WSe$_{2}$\cite{Huang2014,Zhang2015} or MoS$_{2}$-WS$_{2}$\cite{Gong2014}, keeping the metal $M$ fixed and changing the chalcogen $X$ which results in MoS$_{2}$-MoSe$_{2}$ or WS$_{2}$-WSe$_{2}$\cite{Duan2014}, as well as
both changing, as in WSe$_{2}$-MoS$_{2}$\cite{Li2015}.
Recent exciting work has shown remarkable strain control of incommensurate interfaces WSe$_{2}$-MoS$_{2}$\cite{Zhang2018strain} and WS$_{2}$-WSe$_{2}$\cite{Xie2018ParkGroup}, as well as commensurate MoS$_{2}$-WS$_{2}$ and MoSe$_{2}$-WSe$_{2}$\cite{Sahoo2018}, achieving several microns of interfacial length. An atomic sharp interface between two crystalline phases of the same TMD, 1T'-WSe$_{2}$ and 1H-WSe$_{2}$, has also been studied in the context of topologically protected helical edge states\cite{Ugeda2018arxiv}. Doubtlessly, control of lateral HSs in TMDs is being achieved in experiments, and understanding of the interfacial band structure and general behavior is important for future progress.

Theoretical aspects of TMD lateral interfaces have been less studied, especially as one anticipates they could have unique properties, which may have interesting possible uses.
Several works using density functional calculations (DFT) have studied band alignment\cite{Kang2013,Guo2016,OngunOzcelik2016} as well as interface stability and strain\cite{Wei2015,Wei2016,Wei2017}. A HS built of lateral TMD slabs has been predicted to be a high-performance thermoelectric, as the interfaces reduce the thermal conductivity more than electronic mobility\cite{Zhang2016}. Other proposals for lateral HSs include their use as gateless electronic waveguides and spin valley filters/splitters\cite{Ghadiri2018}, as a 1D spin channel driven by just a current flow\cite{Mishra2018oneDimensional}, as well as optoelectronic applications of spatially indirect excitons in these structures \cite{Lau2018Arxiv}.

Motivated by experiments and the unusual nature of lateral HSs,  we study their role in mediating magnetic interactions between impurities at the interface.  To this end, we first model realistic heteroribbons to analyze the main characteristics of these fascinating 1D electronic states. We use a three-orbital tight binding (3OTB) model\cite{Liu2013} that takes DFT and experimental parameters as input. We build real space heterostructures with zigzag and armchair nanoribbons  considering both interfaces and edges, analyzing and contrasting their different characteristics. We find midgap states with clear interfacial and edge character that are highly localized at the corresponding region,
with varying wavefunction orbital and spin content. We also provide an analytical model for the 1D interfacial bands, finding that long-range hoppings up to fourth nearest-neighbor along the interface are important.  This reflects the robustness of the interfacial 1D states, supported by the hexagonal lattice symmetry of TMDs\cite{Segarra2016}, and the hybridization across the interface. Our approach could be used to analyze interfacial states between any two materials or phases of the same TMD, provided that the Hamiltonian for the components is known.

As suspected, we show that the interface provides an effective 1D host with unique characteristics that impacts the physical response of these systems. This is demonstrated by considering the Ruderman-Kittel-Kasuya-Yosida (RKKY) interaction between magnetic impurities on the interface of such lateral HSs\cite{RudermanKittel1954,Kasuya1956,Yosida1957}. It is important to note that magnetic impurities in TMDs are expected to be stable when hybridized in different scenarios\cite{Shao2017,ShiReview2018}. Moreover, substitutional Mn \cite{Zhang2015MagImpurities,Wang2016,Huang2017,Tan2017}, Cr \cite{Huang2017} and Co \cite{Liu2017NatChem,Nethravathi2017} impurities have been recently incorporated into MoS$_2$ flakes. Here we find that the complex spin and orbital texture of the interfacial states results in anisotropic and sizable non-collinear (Dzyaloshinskii-Moriya) effective exchange interactions between the magnetic impurities placed along the interface. The different interaction terms compete with one another and produce unusual ground state alignment of magnetic moments. We further find that this interaction is highly tunable through experimentally accessible parameters, such as gate doping, leading to interaction between impurities which are long ranged, decaying as $\simeq r^{-\frac{1}{2}}$, as the separation $r$ between the impurities increases.

The paper is arranged as follows: In Section \ref{sec:model} we present the tight-binding description for theoretically constructing TMD lateral HSs. In Section\ \ref{sec:EdgeAndInterfaceStates} we analyze the 1D edge states obtained for the zigzag and armchair interfaces. In Section\ \ref{sec:1DhostRKKY} we study the RKKY interaction between magnetic moments at the interface. We give our conclusions in Section\ \ref{sec:conclusions}.

\section{Tight-Binding description}
\label{sec:model}

\begin{figure}[tbph]
\centering
\includegraphics[width=0.5\textwidth]{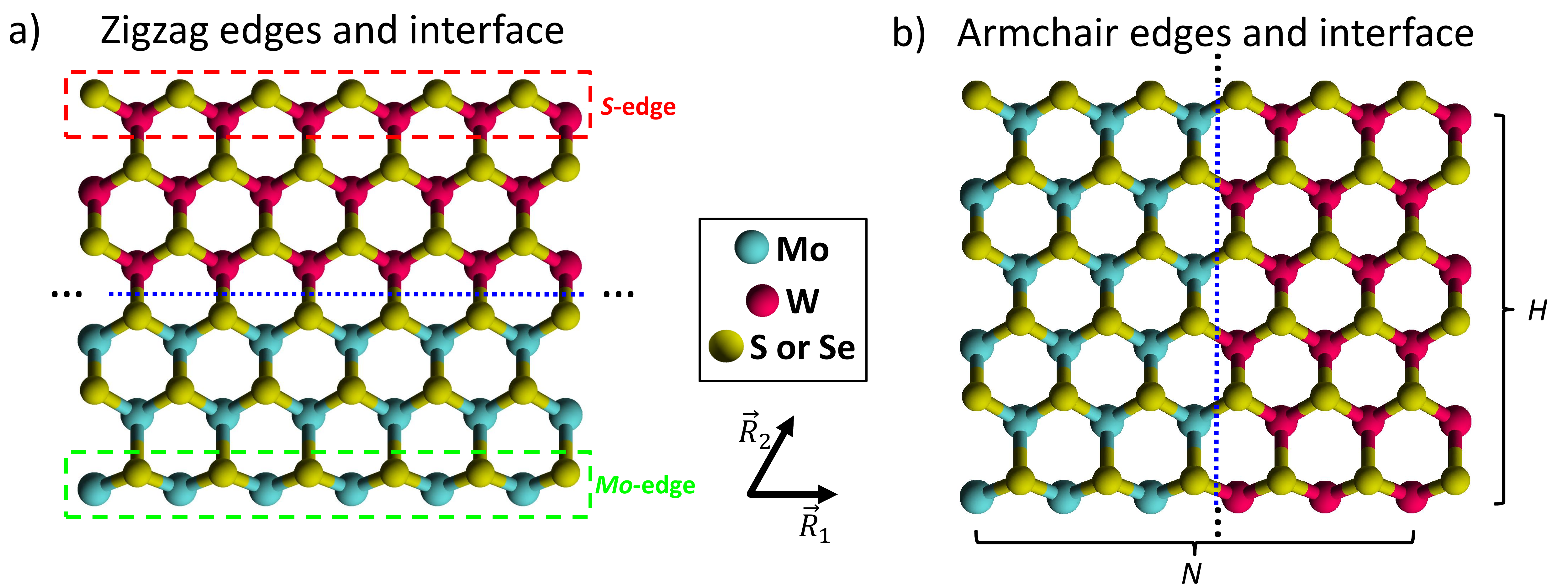}\\
\includegraphics[width=0.5\textwidth]{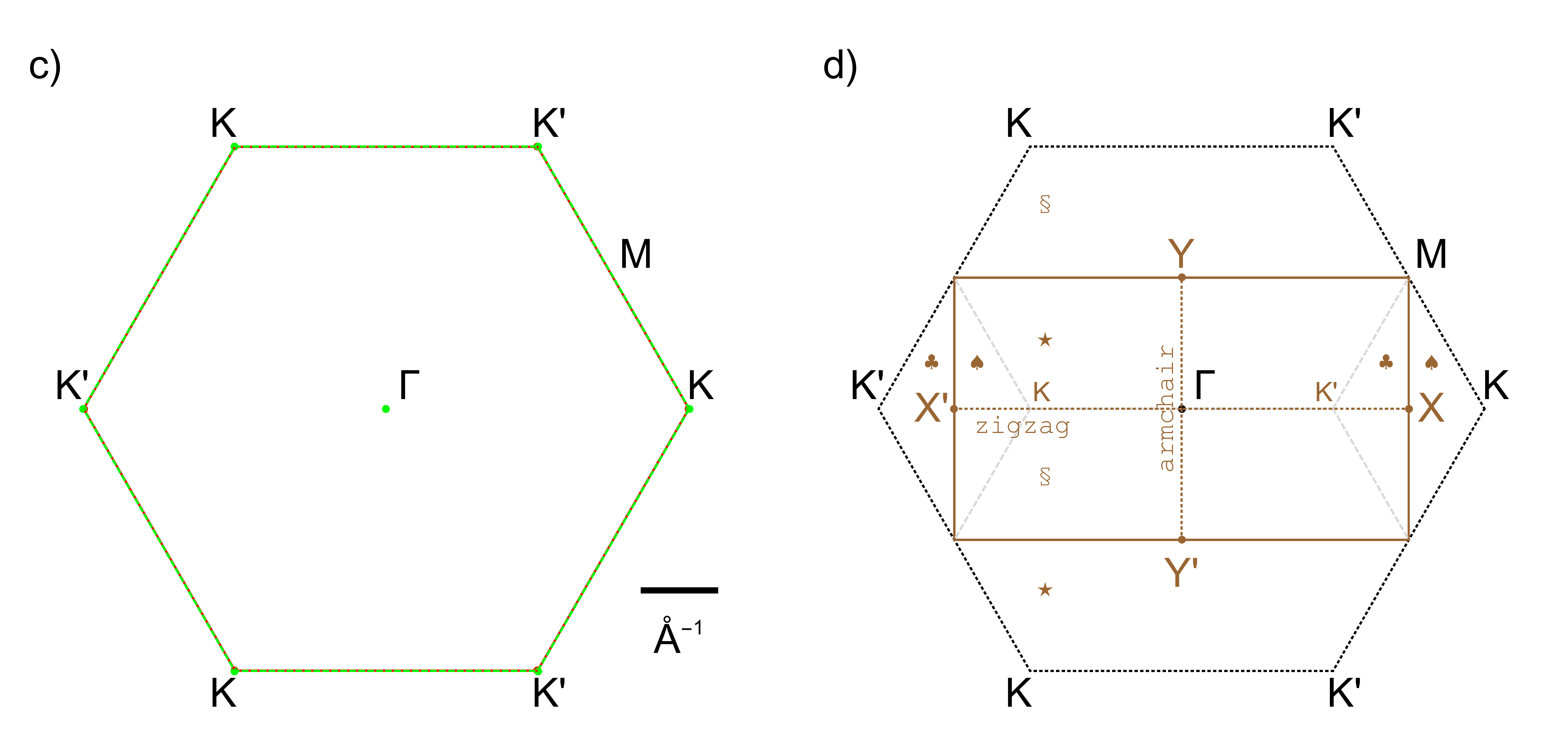}
\caption{Heteroribbons with edges and interfaces for (a) zigzag, and (b) armchair configurations. Metals Mo and W are shown in aqua and red colors, respectively. Chalcogens S or Se are shown in dark yellow. The heteroribbon is finite along one direction, while we use periodic boundary conditions in the other direction, as indicated by the triple black dots. The interface is shown as a blue dotted line. In (a) two different edges present in the zigzag ribbon are shown, the $S$-edge (outermost-atom is a chalcogen) and the $M$-edge (outermost-atom is a transition metal). (c) Brillouin zones for MoS$_2$ (green) and WS$_2$ (red), where differences are noticeable only at the m\AA$^{-1}$ scale. (d) Reduced BZ (brown) for the ribbon as compared to the original BZ (dashed black). The symbols indicate regions in the original BZ, as folded to the reduced BZ.}
\label{Fig1}
\end{figure}

We study commensurate lateral heteroribbons\cite{Huang2014} with realistic sharp interfaces, considering different boundary geometries of edges and interfaces (either zigzag or armchair), with periodic boundary conditions (PBC) along the ribbon. The ribbon can be described by a triangular lattice of metal atoms and associated chalcogens as shown in Fig.\ \ref{Fig1}, with only three 4\emph{d}-orbitals per metal site. This model has been very successful in describing real-space finite structures, such as flakes\cite{Pavlovic2015,AvalosOvando2016PRB,Pawlowski2018} and ribbons\cite{Liu2013,Chu2014,Li2016PRB,Rostami2016,Cortes2018arxiv}, and exploits the fact that the near-gap (low energy) level structure in TMDs is dominated by the metal 4\emph{d}-orbitals with nearly no contribution from the chalcogen \emph{p}-orbitals\cite{Liu2013}. This 3-orbital tight-binding (3OTB) model uses $d_{z^2}$, $d_{xy}$ and $d_{x^2-y^2}$ as basis, and is given in our case by
\begin{equation}\label{heterolattice1}
  H_{\text{3OTB}} = H^{A}_{\text{pristine}} + H^{B}_{\text{pristine}} + H_{\text{interface}},
\end{equation}
where $H^{A(B)}_{\text{pristine}}$ is the Hamiltonian of the two different TMDs, A and B, and $H_{\text{interface}}$ describes the hoppings at the interface between the two TMD lattices. Here, we consider TMDs with the same type of chalcogen atoms, since the lattice mismatch for those structures is less than 1\% (such as MoS$_{2}$-WS$_{2}$ and MoSe$_{2}$-WSe$_{2}$)\cite{Huang2014,Gong2014,Sahoo2018}. This small mismatch  results in corresponding small strain, so that the interface is essentially only compositional. The tight-binding allows one to simply connect the metal atoms across the interface in a one-to-one basis. In contrast, when the chalcogen between $A$ and $B$ is different, the lattice mismatch is about 4\%,\cite{Duan2014} which introduces sizable strain and requires consideration of lattice relaxation effects. Differences in real space lattice sizes are translated into different monolayer Brillouin zones (BZ), as shown in Fig.\ \ref{Fig1}(c), although the difference is in the m\AA$^{-1}$ range and will be neglected. For each of the pristine TMD lattices (A and B), the 3OTB model is given by\cite{Liu2013}
\begin{equation}\label{lattice1}
  H_{\text{pristine}}^{\text{A(B)}} = H^{\text{A(B)}}_{\text{o}} + H^{\text{A(B)}}_{\text{t}} + H^{\text{A(B)}}_{\text{SOC}},
\end{equation}
where $H^{\text{A(B)}}_{\text{o}}$ is the onsite Hamiltonian and $H^{\text{A(B)}}_{\text{t}}$ has the hopping integrals. For each TMD, $H_{\text o}$ is given by
\begin{equation}\label{lattice2}
  H^{\text{A(B)}}_{\text{o}} = \sum_{ \textbf{l}}^{N_{sites}} \sum_{s=\uparrow,\downarrow}^{\text{spin}} \sum_{\alpha,\alpha'}^{\text{orbitals}} \varepsilon^{\text{A(B)}}_{\alpha\alpha',s}d_{\alpha,\textbf{l},s}^{\dagger\text{A(B)}}d^{\text{A(B)}}_{\alpha',\textbf{l},s},
\end{equation}
where $d^{\text{A(B)}}_{\alpha,\textbf{l},s}$ ($d^{\dagger\text{A(B)}}_{\alpha,\textbf{l},s}$) annihilates (creates) a spin-$s$ electron in orbital $\alpha$, $\in\,\left\{d_{z^2},d_{xy},d_{x^2-y^2}\right\}$ in site $\textbf{l}=l_{1}\textbf{R}_{1}+l_{2}\textbf{R}_{2}$, where \textbf{R}$_{j}$ are the lattice vectors of the triangular lattice for each material \cite{supplemental}. For a rectangular ribbon, the total number of sites is $N_{sites}=N\times H$, as shown in Fig.\ \ref{Fig1}(a) and (b). The onsite energies are given by $\varepsilon^{\text{A(B)}}_{\alpha\alpha',s}$, while the nearest-neighbor coupling Hamiltonian is
\begin{equation}
H^{\text{A(B)}}_{\text{t}} = \sum_{\textbf{l,R}_j} \sum_{s=\uparrow,\downarrow} \sum_{\alpha,\alpha'}
t_{\alpha\alpha'}^{(\textbf{R}_{j})\text{A(B)}}d_{\alpha,\textbf{l},s}^{\dagger\text{A(B)}}d^{\text{A(B)}}_{\alpha',\textbf{l}+\textbf{R}_{j},s}+
\text{H.c.},\\
\end{equation}
where $t_{\alpha\alpha'}^{(\textbf{R}_{j})\text{A(B)}}$ are the different hopping parameters, and H.c. is Hermitian conjugate.

The SOC in each material is approximated by the metal onsite contributions, $H^{\text{A(B)}}_{\mathrm{SOC}}=\lambda^{\text{A(B)}} L_{z}S_{z}$, where $L_{z}$ and $S_{z}$ are the $z$-components of the orbital and spin operators, respectively, and $\lambda^{\text{A(B)}}$ is the SOC strength for each material. This results in on-site orbital mixings, $\varepsilon_{d_{xy}d_{x^2-y^2},\uparrow}=\varepsilon_{d_{x^2-y^2}d_{xy},\downarrow}=i\lambda^{\text{A(B)}}$ and $\varepsilon_{d_{xy}d_{x^2-y^2},\downarrow}=\varepsilon_{d_{x^2-y^2}d_{xy},\uparrow}=-i\lambda^{\text{A(B)}}$, that reproduce well the spin-split valence bands in the 2D crystal and give rise to the strong spin-valley locking\cite{Liu2013}. We use $2\lambda^{\text{MoS$_{2}$}}=150$ meV, taken from DFT calculations,\cite{Xiao2012,Liu2013} in good agreement with experiments (152 meV \cite{Sun2013} and $145$ meV \cite{Miwa2015}), while $2\lambda^{\text{WS$_{2}$}}=430$ meV taken from DFT\cite{Xiao2012,Liu2013} and in agreement with experiment (420 meV).\cite{Tanabe2016}

The interface Hamiltonian $H_{\text{interface}}$ is described by nearest neighbor hopping integrals, and needs to take into account two important issues: the band offset (or alignment) between materials $V_{\text{A-B}}$, and rescaling of the hoppings across the interface. The band offset is taken into account through relative shifts of the onsite terms, given by $\varepsilon^{\text{B}'}_{\alpha\alpha',s}=\varepsilon^{\text{B}}_{\alpha\alpha',s}+V_{\text{A-B}}$. These offsets are taken from DFT results\cite{Kang2013} and can result in either type-I or type-II band alignments in these lateral HS. The hopping Hamiltonian connecting the two materials can be written as
\begin{equation}\label{df}
  H_{t}^{\text{A-B}}=\sum_{\textbf{$\gamma$,a}_j} \sum_{s,\alpha,\alpha'}
\delta\left[t_{\alpha\alpha'}^{(\textbf{a}_{j})\text{A}}+t_{\alpha\alpha'}^{(\textbf{a}_{j})\text{B}}\right]d_{\alpha,\textbf{$\gamma$},s}^{\dagger}d_{\alpha',\textbf{$\gamma$}+\textbf{a}_{j},s}+
\text{H.c.},\\
\end{equation}
where $\gamma$ are the atoms on both sides of the interface. $\delta$ is a  scaling factor that describes the compositional symmetry as well as possible relaxation effects at the interface. In what follows we use $\delta=0.1$, which leads to localized states at the interface. Larger $\delta$ produces increasingly delocalized states, as we will explain later, but with similar orbital and spin content features.\cite{supplemental} A geometric average $t^{\text{A-B}}=\sqrt{t^{\text{A}} t^{\text{B}}}$ for commensurate HSs has been used in the literature with similar results.\cite{Zhang2016}  For
non-commensurate HSs, strain can be strong and an averaged $\delta=0.5$ appears to provide a good tight binding description.\cite{Choukroun2018arxiv}

Notice that the heteroribbon naturally yields a band structure in a reduced Brillouin zone (rBZ) instead of the original lattice BZ. The resulting band structure within the rBZ is projected along the 1D-long direction: $N$ for the zigzag, $H$ for the armchair-edge ribbon, see Fig.\ \ref{Fig1}. Similar band projection analysis has been used to study edge states in graphene grain boundaries.\cite{Phillips2015} For the zigzag case, the bands are folded along the horizontal $k_x$-axis in Fig.\ \ref{Fig1}(d), with a $X'X=2\pi/a$ in length. Along this $X'-K-\Gamma-K'-X$ path, the valleys are still decoupled after band folding, with $K$ ($K'$) located at $-2\pi/3a$ ($2\pi/3a$). For the armchair case, the band projection along the vertical $k_y$-axis in Fig.\ \ref{Fig1}(d) creates a rBZ with shorter length $Y'Y=2\sqrt{3}\pi/3a$ (as compared to the zigzag ribbon), reflecting the larger armchair unit cell\cite{supplemental}. This band folding overlaps the $K$ and $K'$ valleys with $\Gamma$.\cite{Peterfalvi2015,Rostami2016,Chen2017armmchair,Davelou2017} In addition to the band folding, edge states will appear within the band gap in the nanoribbon HS. We will see this behavior in the rBZ for MoS$_2$-WS$_2$ and MoSe$_2$-WSe$_2$ systems in the following section.

\section{Edge and interface states}
\label{sec:EdgeAndInterfaceStates}

As described in Fig.\ \ref{Fig1}, we consider heteroribbons of MoS$_{2}$-WS$_{2}$, with either zigzag termination and interface, as depicted in Fig.\ \ref{Fig1}(a), or with armchair edges and interface, in Fig.\ \ref{Fig1}(b)\cite{supplemental}. Both types of heterojunctions have been seen experimentally, although zigzag\cite{Huang2014,Duan2014,Gong2014,Sahoo2018} is more recurrent than armchair termination\cite{Gong2014}.
We consider heteroribbons with $N=100$ and $H=40$ typical for a heterostructure of 4,000 metal atoms. This size is found to be sufficiently large to clearly identify localized wave functions at either the edges or interface, without cross-interference\cite{AvalosOvando2016PRB,Wei2017}. These sizes correspond to $\sim 350 \text{nm}^2$, comparable to experimentally available interfaces in heterotriangles\cite{Huang2014,Duan2014,Gong2014}, or ribbons with straight edges\cite{Cheng2017NanoLett,Chen2017NatComm,Zhao2018NanoLett,Cui2018NatComm}. We consider a type-II band alignment of $V_{\text{MoS$_2$-WS$_2$}}=0.242$ eV as proposed from DFT calculations\cite{Kang2013} and confirmed experimentally with scanning tunneling spectroscopy\cite{Hill2016}, and a combination of ultraviolet and X-ray photoelectron spectroscopy\cite{Chiu2017} in MoS$_{2}$-WS$_{2}$ vertical heterostructures. For the selenium-based HS we use $V_{\text{MoSe$_2$-WSe$_2$}}=0.262$ eV.\cite{Kang2013}

We are specially interested in the interface states located at the junction between both materials. We numerically diagonalize the full Hamiltonian and Fourier transform the states to extract the respective momenta and build the projected band structure shown in Figs.\ \ref{Fig2} and \ref{Fig5}. Spin up (down) states are shown as black (gray) dots, demonstrating time reversal symmetry of the spectrum, as $\tau \varphi(k,s)=\varphi(-k,-s)$ and $\varphi(k,s)$ are degenerate. Most notably, in addition to the typical bulk bands, there are states crossing the gap similar to those obtained in the direct $k$-space continuum solution of the 2D bulk tight-binding Hamiltonian [see Appendix A in Ref.\ \onlinecite{Liu2013}]. The bands dispersing across the gap can be seen to be located at either the edges or at the interface of the nanoribbon, with at least 90\% of their spatial weight at the corresponding atomic rows in either MoS$_{2}$ or WS$_{2}$ edges, or at the interfacial region between both materials. Figures\ \ref{Fig2} and \ref{Fig5} label the states with different color symbols depending on their locations: green if located at the edges of MoS$_{2}$, red if in WS$_{2}$, and blue when at the interface MoS$_{2}$-WS$_{2}$; the symbol size reflects the wave function magnitude squared.

The model allows one to identify the real-space location of the midgap energy states and could be used to introduce defects, such as vacancies and adatoms. We should mention that such defects have been shown to produce only slight deviations from this pristine band structure\cite{Chu2014}, in addition to creating midgap localized states\cite{Wei2016}.

\subsection{Zigzag interface states}
\label{subsec:ZigzagInterfaceStates}

\begin{figure*}[tbph!]
\centering
\includegraphics[width=0.33\textwidth]{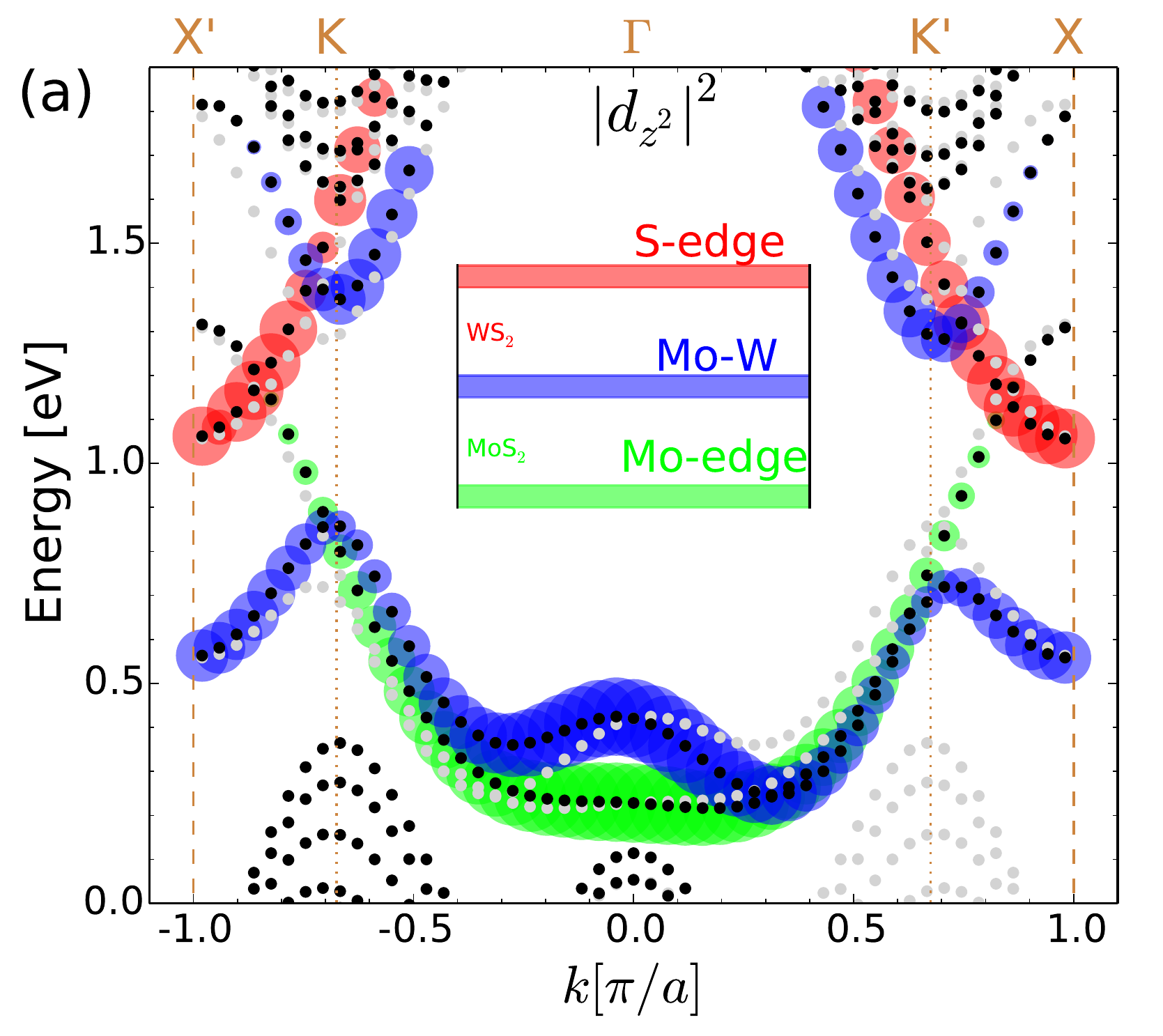}\includegraphics[width=0.33\textwidth]{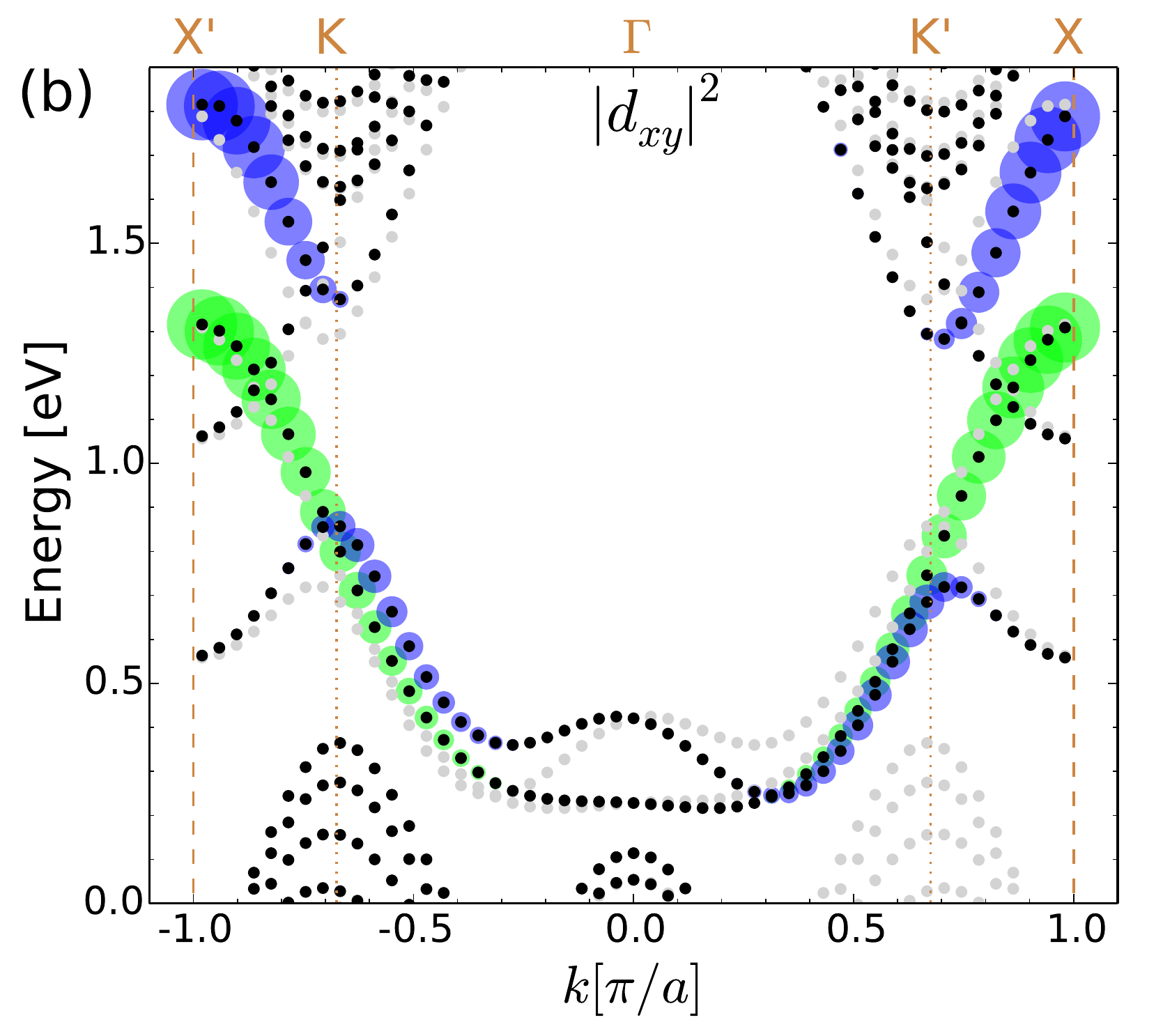}\includegraphics[width=0.33\textwidth]{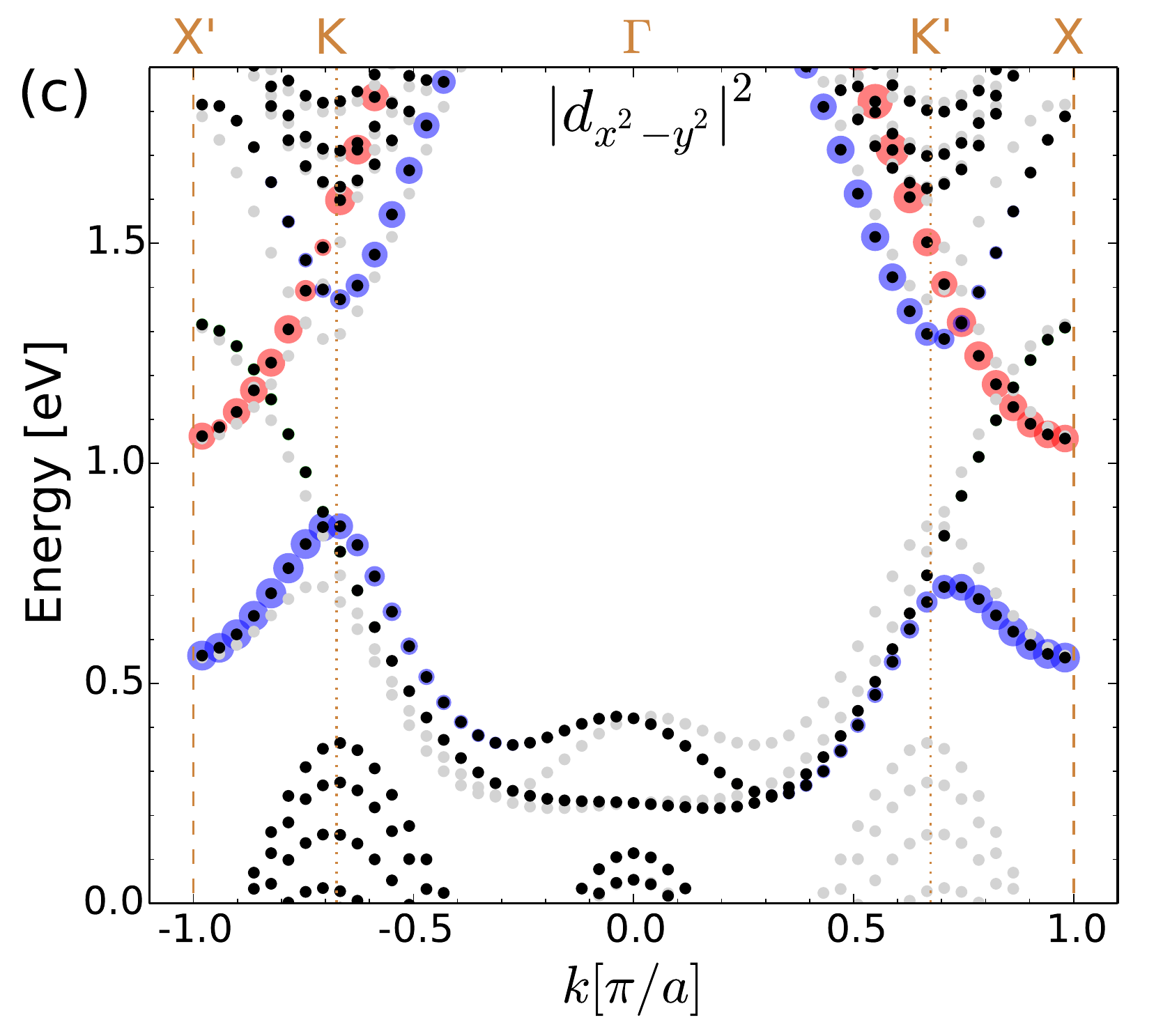}\\
\caption{Band dispersions for the zigzag edge/interface HS ribbon as shown in Fig.\ \ref{Fig1}(a) projected along $X'-K-\Gamma-K'-X$ direction. Spin up (down) states are shown in black (gray) dots. The orbital wave function weights (a) $|d_{z^2,\text{s}}|^2$, (b) $|d_{xy,\text{s}}|^2$, and (c) $|d_{x^2-y^2,\text{s}}|^2$ are proportional to the size of colored circles, with different color indicating the location of the wave function throughout the ribbon: Mo-edge (green), W-edge (red) and interface (blue), as schematized in the insets in (a); only shown for spin up for clarity. Selected states show at least 90\% of the wave function at either of the three locations shown at the insets.}
\label{Fig2}
\end{figure*}

First, let us briefly discuss the edge states found in the pristine MoS$_{2}$ and WS$_{2}$ sides of the ribbon. As mentioned before, in the single material zigzag ribbon, one can find two distinct edges: the $M$-edge and the $S$-edge. In the first one, the outermost atom is a transition metal, while in the latter the outermost atom is a chalcogen, as schematically shown in Fig.\ \ref{Fig1}(a). In general, one can find highly localized states at either edge, with opposite dispersion across the gap. The $M$-edge ($S$-edge) band has positive (negative) mass around the $\Gamma$ point, which reverses sign at $X$ or $X'$. In Fig.\ \ref{Fig1}(a) one can see the $M$-edge of the MoS$_{2}$ ribbon (green symbols) and the $S$-edge of the WS$_{2}$ side (red symbols), since those edges remain pristine after the interface is formed.\footnote{Two other edge states arising from higher energy orbital bands in the bulk are not captured within the 3OTB model\cite{Liu2013}, but are not expected to cross the bulk band significantly.} These states lie on the outer edges of the ribbon and can be seen closing the gap in Fig.\ \ref{Fig2}(a-c). The various orbital weights are qualitatively the same as for a single TMD zigzag ribbon\cite{supplemental}. This is in good agreement with other theoretical continuous-$k$ approaches within the same model\cite{Liu2013,Chu2014,Li2016PRB,Rostami2016}.

Let us now discuss the hybridization at the interface of the zigzag heteroribbon. As shown in Fig.\ \ref{Fig2}(a-c), there are two interfacial midgap bands (blue), one closer to the conduction band and other to the valence band, which we are going to call upper interfacial band (UIB) and lower interfacial band (LIB), respectively. These UIB and LIB have different weights in all three orbital components $d_{z^2,\text{s}}$, $d_{xy,\text{s}}$ and $d_{x^2-y^2,\text{s}}$. The gap between these two branches is proportional to the value of the hybridization parameter $\delta$, and is the result of the hybridization of the $S$-edge in the MoS$_{2}$-side and the $M$-edge in the WS$_{2}$ side\cite{supplemental}. The hopping integrals that mediate this edge hybridization then produce coherent 1D states that have sizable amplitudes on {\em both} sides of the HS.\@ As such, they carry information on orbital and spin components, as well as the relative band offsets that impart them with interesting properties, as we will discuss further. In general, the orbital weights follow $|d_{z^2,\text{s}}|^2 \simeq 2|d_{xy,\text{s}}|^2 \simeq 2|d_{x^2-y^2,\text{s}}|^2$, as qualitatively seen in clear pristine zigzag terminations\cite{Pavlovic2015,Segarra2016}. In particular, the LIB shows more MoS$_2$ weight at the edges of the zone ($k\rightarrow \pm\pi/a$) and more WS$_2$ weight in the middle of the zone ($k\rightarrow 0$), while for the valley projections ($k=-2\pi/3a$ for K and $k=2\pi/3a$ for K') the wave function is equally distributed among both materials. No interface edge states are found for the orbital $d_{xy,\text{s}}$ on the Mo side, or $d_{x^{2}-y^{2},\text{s}}$ on the W side, for either spin, due to the orbital symmetry of the zigzag terminations\cite{Pavlovic2015}.

\begin{figure}[tbph]
\centering
\includegraphics[width=0.45\textwidth]{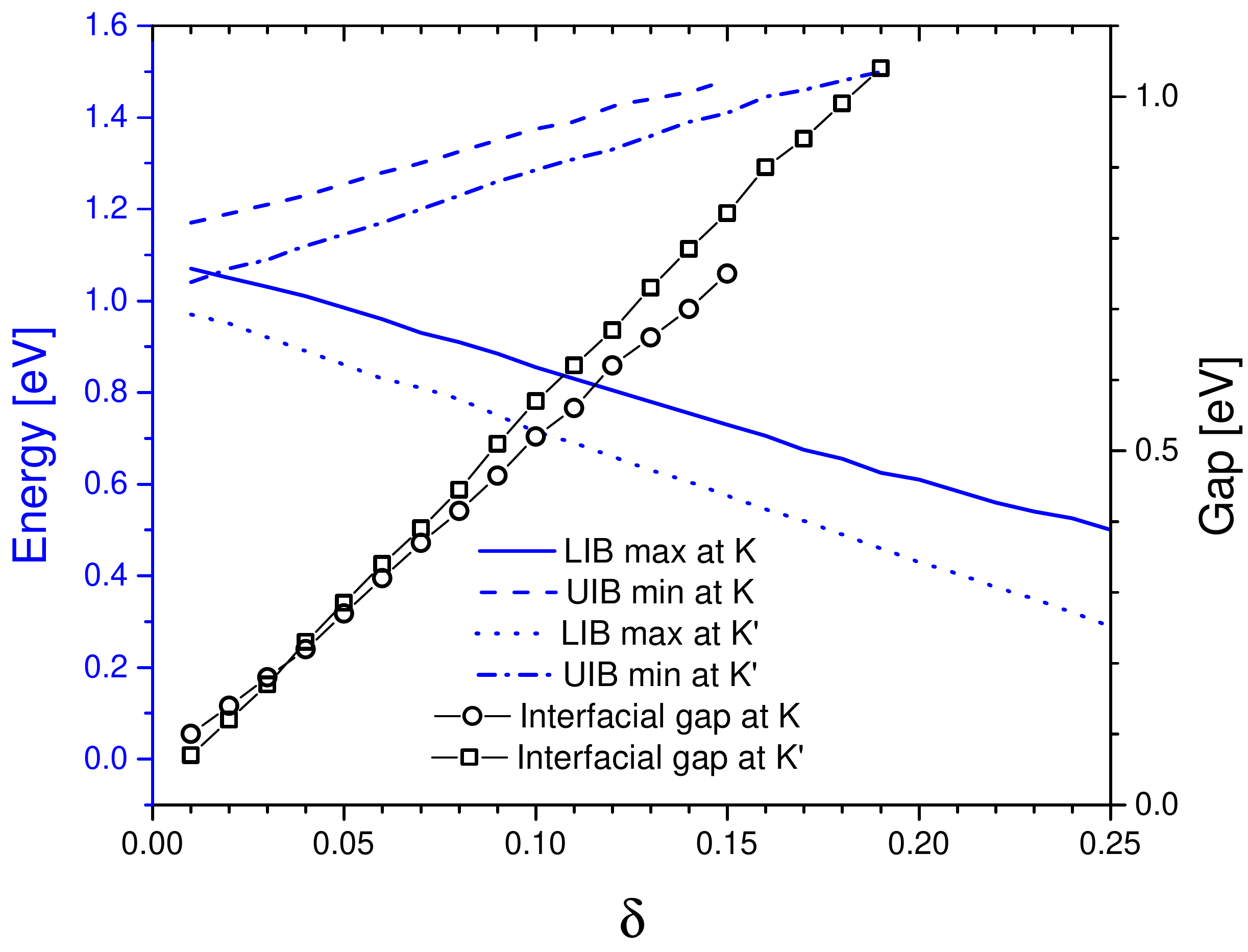}\\
\caption{$\delta$-dependence of interface states for a zigzag MoS$_2$-WS$_2$ HS, for spin up states. In blue: maxima and minima for edge lower interfacial band (LIB) and upper interfacial band (UIB), respectively, at each valley projection. In black: interfacial gap between the UIB minima and LIB maxima at each valley. The gap is nearly linear in $\delta$.}
\label{Fig3}
\end{figure}

As the structure of the HS depends on the hybridization between both materials, let us comment further on the effect of the contrast in hopping integrals parameterized by $\delta$\cite{supplemental}. The HS hybridization naturally creates a gap in the zigzag interfacial states, linearly proportional to $\delta$ for small $\delta$, as shown in black symbols in Fig.\ \ref{Fig3}, while for larger $\delta$, the interface states become fully hybridized to the bulk bands. The interfacial gap increases linearly from zero when $\delta=0$ to $\approx 1.0$ eV ($\approx 0.75$ eV) when $\delta=0.19$ ($\delta=0.15$) for the $K'$ ($K$) valley. Notice that $\delta\rightarrow0$ would recover the natural heteroribbon behavior, closing the gap in the zigzag case. For larger $\delta$, the interfacial bands hybridize with bulk bands, as the band offset and similarity in hoppings across the HS produce a nearly transparent interface. This is shown in Fig.\ \ref{Fig3} with blue lines, where we follow the maxima and minima of the LIB and UIB around each valley, respectively, until they become untraceable due to the hybridization to the bulk bands. The hybridization decreases for $\delta\rightarrow0$ (with metallic behavior for zigzag), while for $\delta \gtrsim 0.4$ the edge bands are fully hybridized to the bulk.  The LIB is still visible at $\delta=0.3$, with the UIB visible at $\delta=0.2$. This behavior holds for different heteroribbon sizes, as well as for the MoSe$_2$-WSe$_2$ heterostructure (not shown).


\begin{figure}[tbph]
\centering
\includegraphics[width=0.45\textwidth]{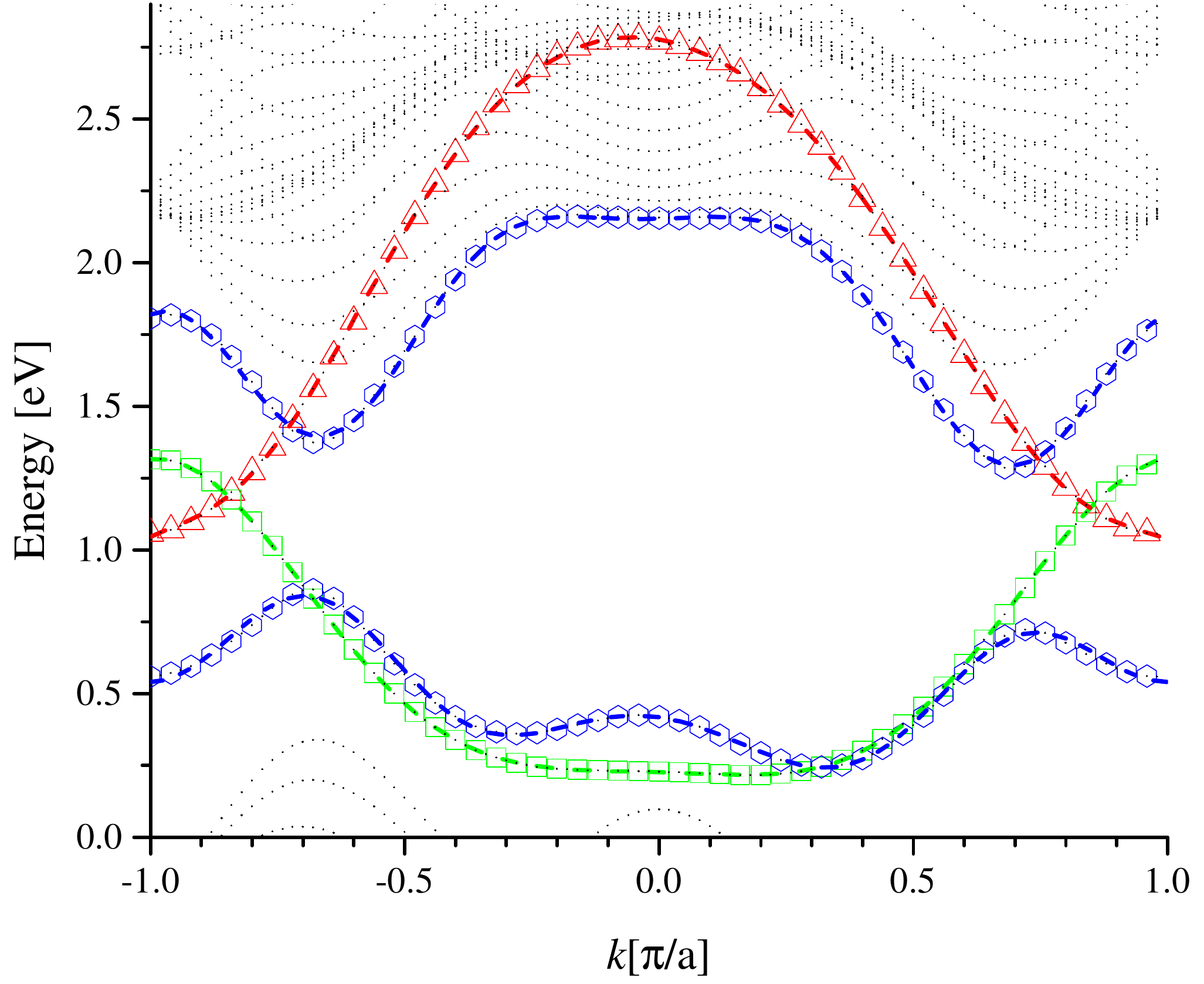}\\
\caption{Fitted bands for the zigzag MoS$_2$-WS$_2$ heteroribbon: The fits to Eq.\ \ref{EffectiveHamiltonianWithPauli} are shown as dashed lines, while symbols indicate the numerical 3OTB bands as shown in Fig.\ \ref{Fig2}(a-c). We highlight interfacial zigzag bands (blue hexagons), as well as pristine Mo (green squares) and W (red triangles) edge bands. Only spin up states are shown. Parameters of the fit are given in Table\ \ref{tab:tablefitting}.}
\label{Fig4}
\end{figure}

\begin{table*}[tbph!]
\caption{\label{tab:tablefitting}%
Fitted parameters of Eq.\ \ref{EffectiveHamiltonianWithPauli} for midgap LIB [$t^{(n)}$ and $t^{(n)}_{SO}$] and UIB [$\gamma^{(n)}$ and $\gamma^{(n)}_{SO}$] of wave functions located on both sides of the heteroribbon interface. Single $t$'s and $\gamma$'s are in eV. Fit parameters for ribbon outer pristine edges not listed.
}
\begin{ruledtabular}
\begin{tabular}{ccccccccccccccccccc}
 & $t^{(0)}$ & $t^{(1)}$ & $\frac{t^{(1)}}{t^{(2)}}$ & $\frac{t^{(1)}}{t^{(3)}}$ & $\frac{t^{(1)}}{t^{(4)}}$ & $t_{SO}^{(1)}$ & $\frac{t_{SO}^{(1)}}{t_{SO}^{(2)}}$ & $\frac{t_{SO}^{(1)}}{t_{SO}^{(3)}}$ & $\frac{t_{SO}^{(1)}}{t_{SO}^{(4)}}$ & $\gamma^{(0)}$ & $\gamma^{(1)}$ & $\frac{\gamma^{(1)}}{\gamma^{(2)}}$ & $\frac{\gamma^{(1)}}{\gamma^{(3)}}$ & $\frac{\gamma^{(1)}}{\gamma^{(4)}}$ & $\gamma_{SO}^{(1)}$ & $\frac{\gamma_{SO}^{(1)}}{\gamma_{SO}^{(2)}}$ & $\frac{\gamma_{SO}^{(1)}}{\gamma_{SO}^{(3)}}$ & $\frac{\gamma_{SO}^{(1)}}{\gamma_{SO}^{(4)}}$ \\
\colrule
\\
MoS$_2$-WS$_2$   & 0.511 & -0.19 & 45.3 & -1.5 & 6.9 & -0.08 & -32.6 & -7.5 & 8.9 & 1.793 & 0.34 & 2.1 & -1.9 & 11.7 & -0.04 & -1.5 & 2.1 & -2.6  \\
MoSe$_2$-WSe$_2$ & 1.032 & -0.19 & 2.4 & -1.9 & 8.6 & -0.08 & -30.1 & -7.5 & 9.6 & 2.145 & 0.23 & 1.8 & -1.5 & 7.6 & -0.04 & -1.4 & 2.3 & -3.4  \\
\end{tabular}
\end{ruledtabular}
\end{table*}

While the numerical approach is needed for a full description of the lattice, analytical models provide a complementary and efficient description. Low-energy analytical models for zigzag $M$-edges\cite{Xu2014}, or chalcogen terminated $S$-edges\cite{Chu2014}, have described the valley dynamics with models up to order $k^4$. In the case of HS interfacial 1D states, the dependence is much more complicated.  We propose here an analytical model to describe zigzag HS states.  Considering time reversal but lack of inversion symmetry in TMDs, a 1D effective Hamiltonian can be written as
\begin{eqnarray}\label{EffectiveHamiltonianWithPauli}
H^{\text{interface}}_{\text{eff}} &= \frac{(1-\sigma_{z})}{2} \sum_{n=0}^{\mathcal{N}}\left[t^{(n)}\cos (n k) + s_{z} t^{(n)}_{SO}\sin (n k)\right] + \nonumber\\
&\frac{(1+\sigma_{z})}{2} \sum_{n=0}^{\mathcal{N}}\left[\gamma^{(n)}\cos (n k) + s_{z} \gamma^{(n)}_{SO}\sin (n k)\right],
\end{eqnarray}
where $\sigma_{z}$ is the Pauli matrix in a two function basis $\left\{ |\phi_{c}\rangle , |\phi_{v}\rangle \right\}$ and $s_{z}$ is the corresponding spin operator. The constants are related to the $n$th-nearest neighbor hoppings $t^{(n)}$ ($\gamma^{(n)}$) and to the $n$th-nearest neighbor spin-orbit interaction $t^{(n)}_{SO}$ ($\gamma^{(n)}_{SO}$) for the LIB (UIB) in the gap, respectively. These parameters are obtained by fitting to the 3OTB band structure calculations, and given in Table\ \ref{tab:tablefitting} for zigzag MoS$_2$-WS$_2$ and MoSe$_2$-WSe$_2$ heteroribbons.

One can see in Fig.\ \ref{Fig4} that the interfacial/edge bands fitting is excellent throughout the entire BZ.
We find that long-range hopping interactions ($\mathcal{N}=4$) are needed. The effective dimensionality of the interface is indeed affected by the bulk lattice sites not at the interface, as well as the hybridization across the interface.  A similar result is seen in the SOC hopping parameters, where at least $\mathcal{N}=3$ is needed. The selenium-based heterostructure MoSe$_2$-WSe$_2$ bands are not shown in Fig.\ \ref{Fig4}, but require the same $\mathcal{N}$ values. These results suggest that the influence of distant neighbors is significant, as enhanced by the HS interface, and may be reflected in interesting 1D physics involving these states.

\subsection{Armchair interface states}
\label{subsec:ArmchairInterfaceStates}

\begin{figure*}[tbph!]
\centering
\includegraphics[width=0.395\textwidth]{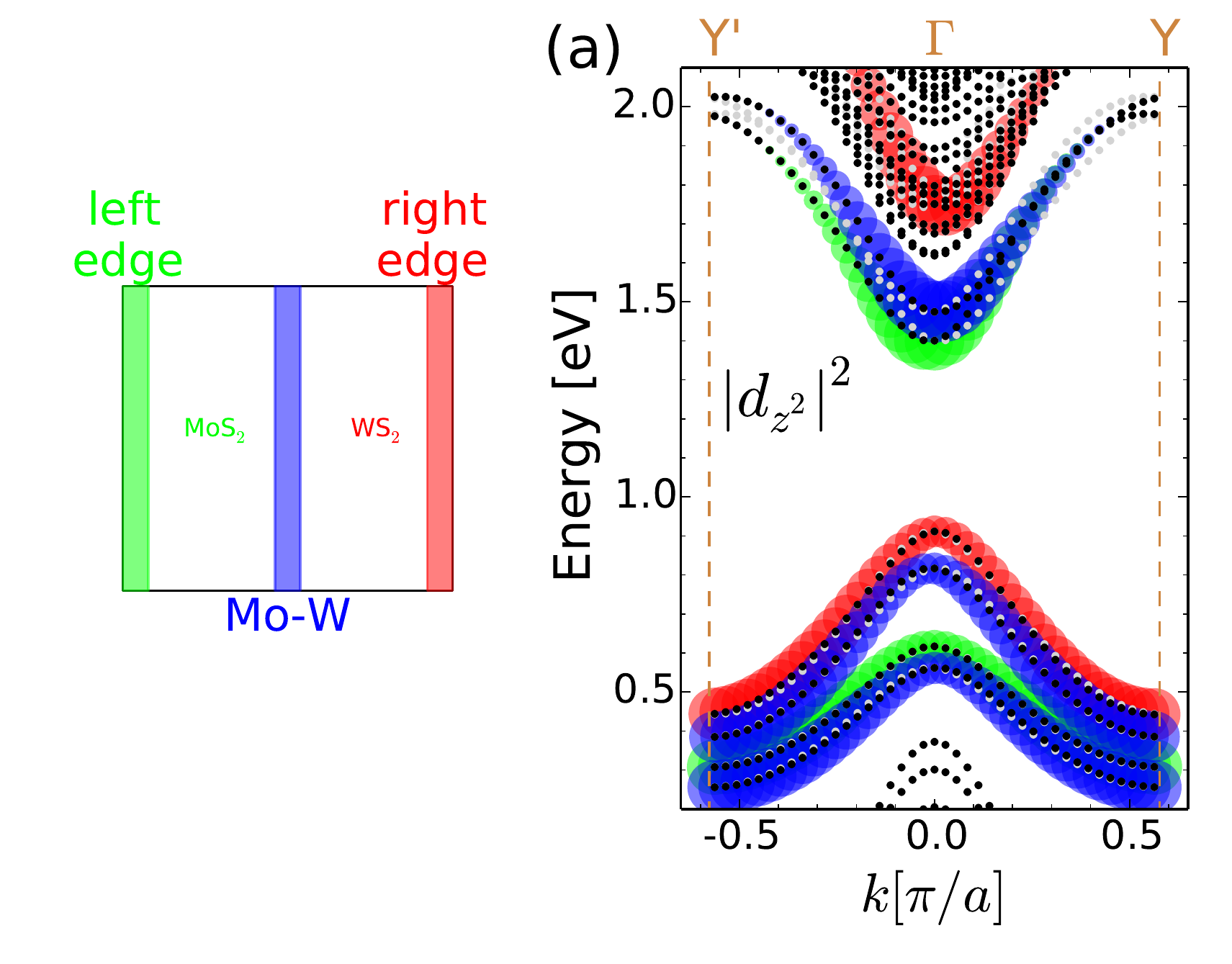}\includegraphics[width=0.327\textwidth]{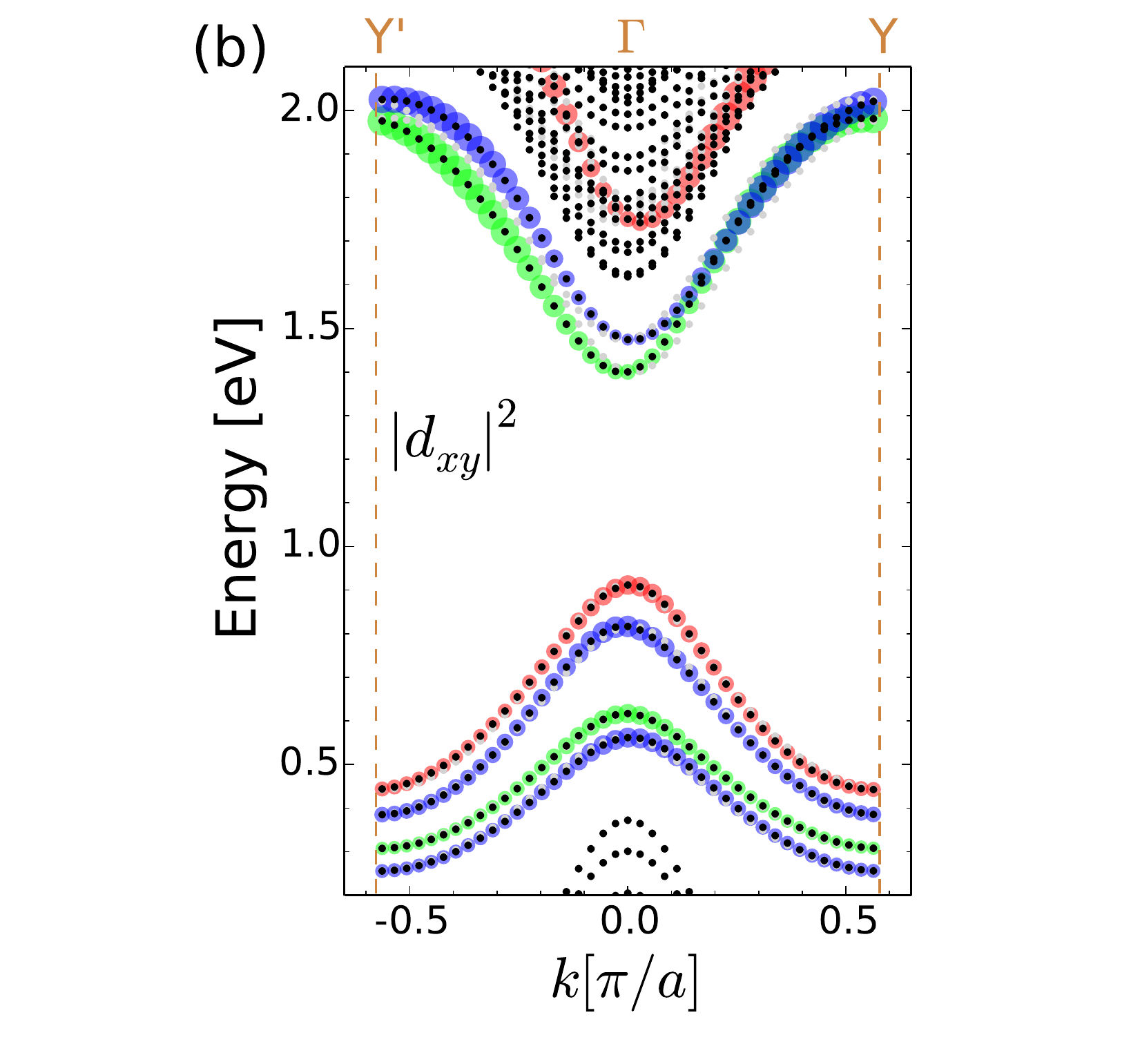}\includegraphics[width=0.365\textwidth]{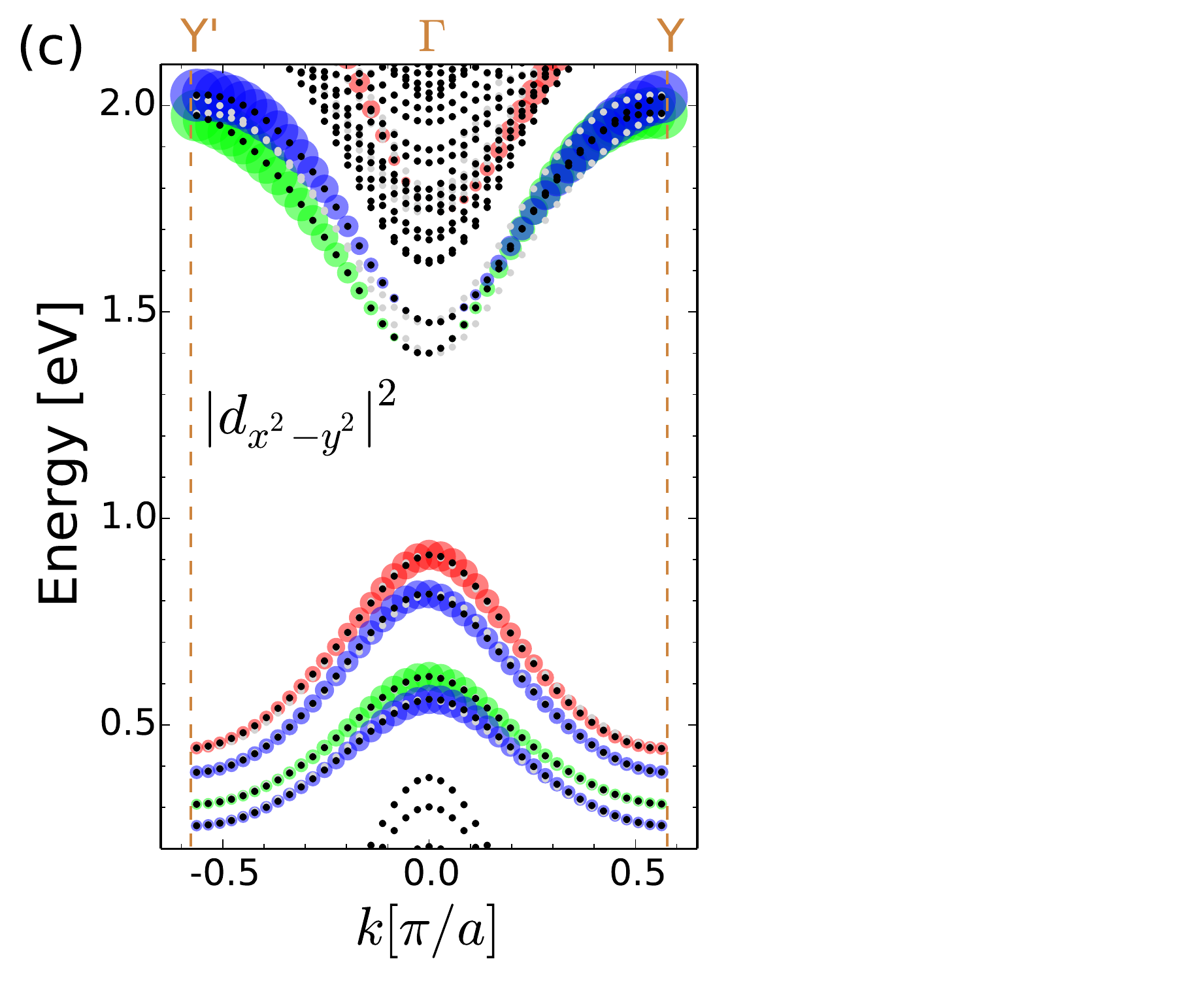}\\
\caption{Band dispersions for the armchair edge/interface HS ribbon as shown in Fig.\ \ref{Fig1}(b) along $Y'-\Gamma-Y$. Description of the curves is the same as in Fig.\ \ref{Fig2}.}
\label{Fig5}
\end{figure*}

The pristine armchair ribbons have not been studied in much detail, perhaps as they are expected to be less stable. Rostami \emph{et al.}\cite{Rostami2016} find gapped edge modes within the reduced BZ, in agreement with DFT calculations\cite{Li2008Ribbons,Ataca2011}, attributed to intervalley scattering from the mixing of 1D-valleys on the edge. As seen in Fig.\ \ref{Fig1}(b), both edges of the armchair ribbon are indistinguishable in a pristine material, leading to the creation of nearly-degenerate gapped edge states lying in the bulk gap\cite{supplemental}. A pair sits close to the valence bulk bands and another pair close to the conduction band\cite{supplemental}. In the heterostructure, Mo-edge and W-edge pristine edge bands can be seen in Fig.\ \ref{Fig5}(a-c) within the gap, in green and red colors, respectively. As before, and unlike the zigzag case, these edge states do not close the heterostructure gap, regardless of the size of the system.

The interfacial states in the armchair heteroribbon are shown in blue in Fig.\ \ref{Fig5}(a-c) for all orbitals. The magnitude of the interfacial wave functions is typically much larger for the $d_{z^{2},\text{s}}$ orbital than for the other two, $|d_{z^2,\text{s}}|^2 \simeq 10|d_{xy,\text{s}}|^2 \simeq 10|d_{x^2-y^2,\text{s}}|^2$. After hybridization in the HS, the type-II alignment allows for easy differentiation of two interfacial bands (blue symbols) in the bulk band gap, one per each material, but displaced to lower energy with respect to the pristine edge band. For bands close to the bulk conduction bands, the hybridization process is similar, with the interface state fully hybridized to the bulk conduction bands, and barely visible. The interfacial band associated to the Mo-side is visible and has been displaced to higher energy.

\begin{figure}[tbph]
\centering
\includegraphics[width=0.45\textwidth]{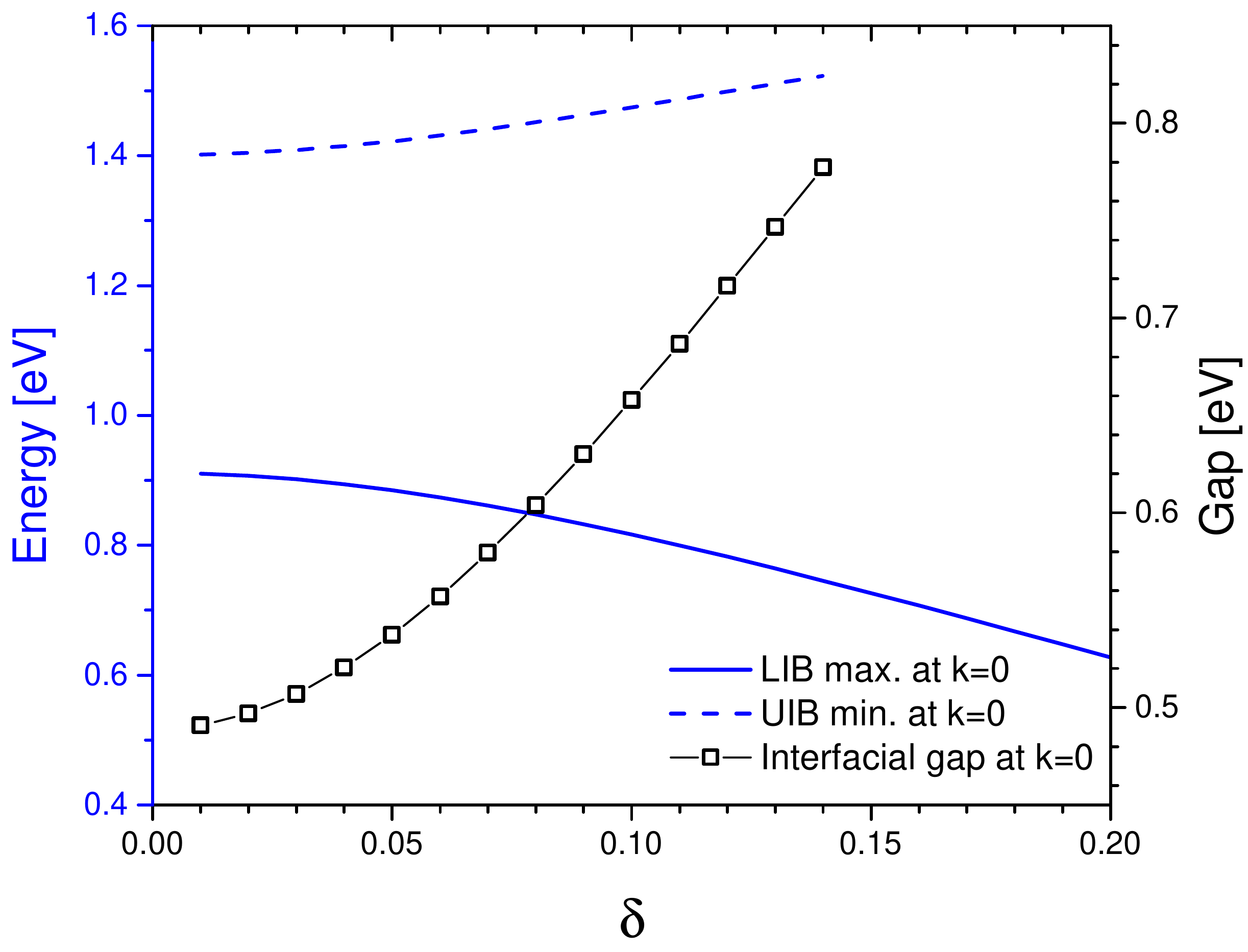}\\
\caption{$\delta$-dependence for an armchair MoS$_2$-WS$_2$ HS, for spin up states. In blue: maxima and minima for edge lower interfacial band (LIB) and upper interfacial band (UIB), respectively, at $k=0$. In black: interfacial gap between the UIB minima and LIB maxima at $k=0$.}
\label{Fig6}
\end{figure}

The gap for the interfacial bands also scales with $\delta$, as in the zigzag case, except that for small $\delta$ the gap does not close, as shown in black symbols Fig.\ \ref{Fig6}. The HS hybridization increases the gap in the armchair interfacial states quadratically proportional to small $\delta<0.07$ (gap $\approx0.58$ eV), and linearly proportional for larger $\delta$, up to a gap of $\approx0.78$ eV when $\delta=0.14$. This difference in the proportionality of the gap with $\delta$ is due to the band folding occurring at $k=0$, where $\Gamma$, $K$ and $K'$ points are contributing. For larger $\delta$, the gap becomes untraceable and fully hybridized to the bulk. At $\delta\rightarrow0$ the armchair semiconducting natural pristine ribbons behavior is recovered, restoring the gapped edge symmetry. For larger $\delta$, the trend is similar as in the zigzag case. The interfacial bands hybridize with bulk bands, reaching a fully hybridization around $\delta \gtrsim 0.4$, product a nearly transparent interface. The lower interfacial band is still visible at $\delta=0.4$ (solid blue line in Fig.\ \ref{Fig6}), with parts of the upper interfacial band are barely visible at $\delta=0.2$, with the minimum already lost in the bulk conduction band (dashed blue line in Fig.\ \ref{Fig6}). The same behavior is observed for the MoSe$_2$-WSe$_2$ heterostructure (not shown). More details of the difference between pristine and HS armchair ribbons and the hybridization of the bands at the interface can be found in the supplement \cite{supplemental}.

\section{1D platform host}
\label{sec:1DhostRKKY}

As suggested above, the interface states in this lateral HS could act as an effective 1D host with interesting properties. We explore here the Ruderman-Kittel-Kasuya-Yosida (RKKY) interaction \cite{RudermanKittel1954,Kasuya1956,Yosida1957} when two magnetic impurities are placed at the interface. In general, pristine TMDs and their HSs could act as suitable platforms for a tunable RKKY interaction, since they can reach conductive character\cite{Lu2014MIDGAP,Ponomarev2018MIDGAP} and provide stable hosts for magnetic impurities\cite{Shao2017,ShiReview2018,Zhang2015MagImpurities,Wang2016,Huang2017,Tan2017,Liu2017NatChem,Nethravathi2017}.

The RKKY interaction is typically a combination of an oscillatory function and an envelope decaying usually with a power related to the dimensionality of the host system, with a prefactor that depends on the density of states at the Fermi level. The interaction can then be written as $\propto \cos{(2 k_F r)}/r^{d}$, where $r$ is the distance between impurities, $k_{F}$ is the Fermi momentum and $d$ is the dimensionality of the host electron system. The oscillatory term $\cos{(2 k_F r)}$ changes the character of the interaction between ferromagnetic (FM) and antiferromagnetic (AFM) alignment. On the other hand, the decaying envelope limits how far the impurities can see each other. In 2D metallic systems $d=2$, i.e.\ the interaction decays quadratically, and essentially vanishes when the impurities are just a few sites apart. Glimpses of a peculiar \emph{sub-2D} behavior have been spotted in some systems such as edges in TMDs \cite{AvalosOvandoJPCM2018}($d \simeq 1/2$), graphene \cite{Black2010,Duffy2014} ($1<d<2$), and silicine \cite{Zare2016}($d \simeq 1$), usually driven by orbital content and symmetry of the host state electrons mediating the interaction.
It is interesting that recent work finds additional non-decaying interaction terms for 1D quantum wires with SOC, which would appear as producing $d \lesssim 1$ behavior \cite{Silva2018unpublish}.
We anticipate that the unusual behavior of the HS interface states discussed above would fall in this general group as well.

\begin{figure}[tbph]
\centering
\includegraphics[width=0.45\textwidth]{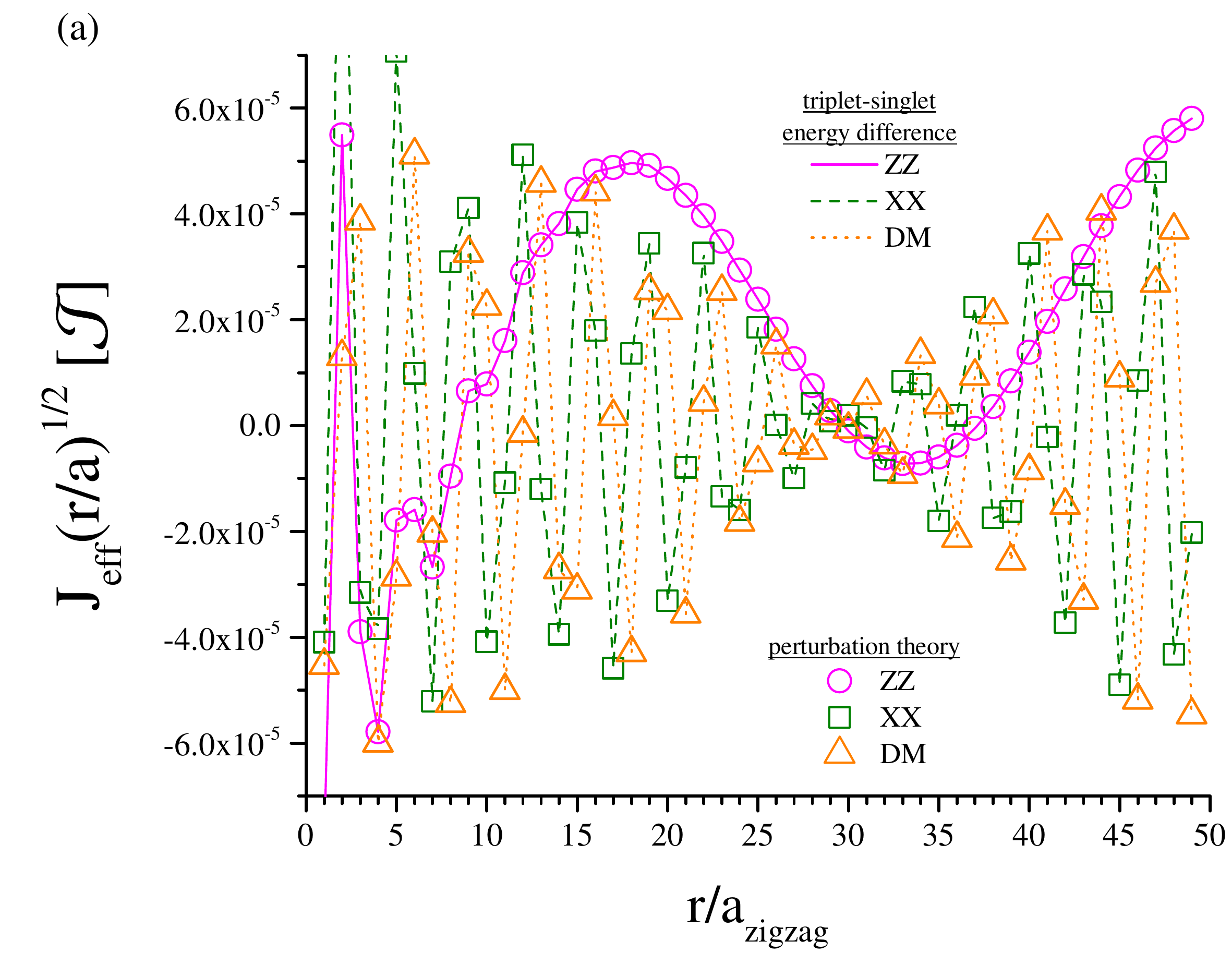}\\
\includegraphics[width=0.45\textwidth]{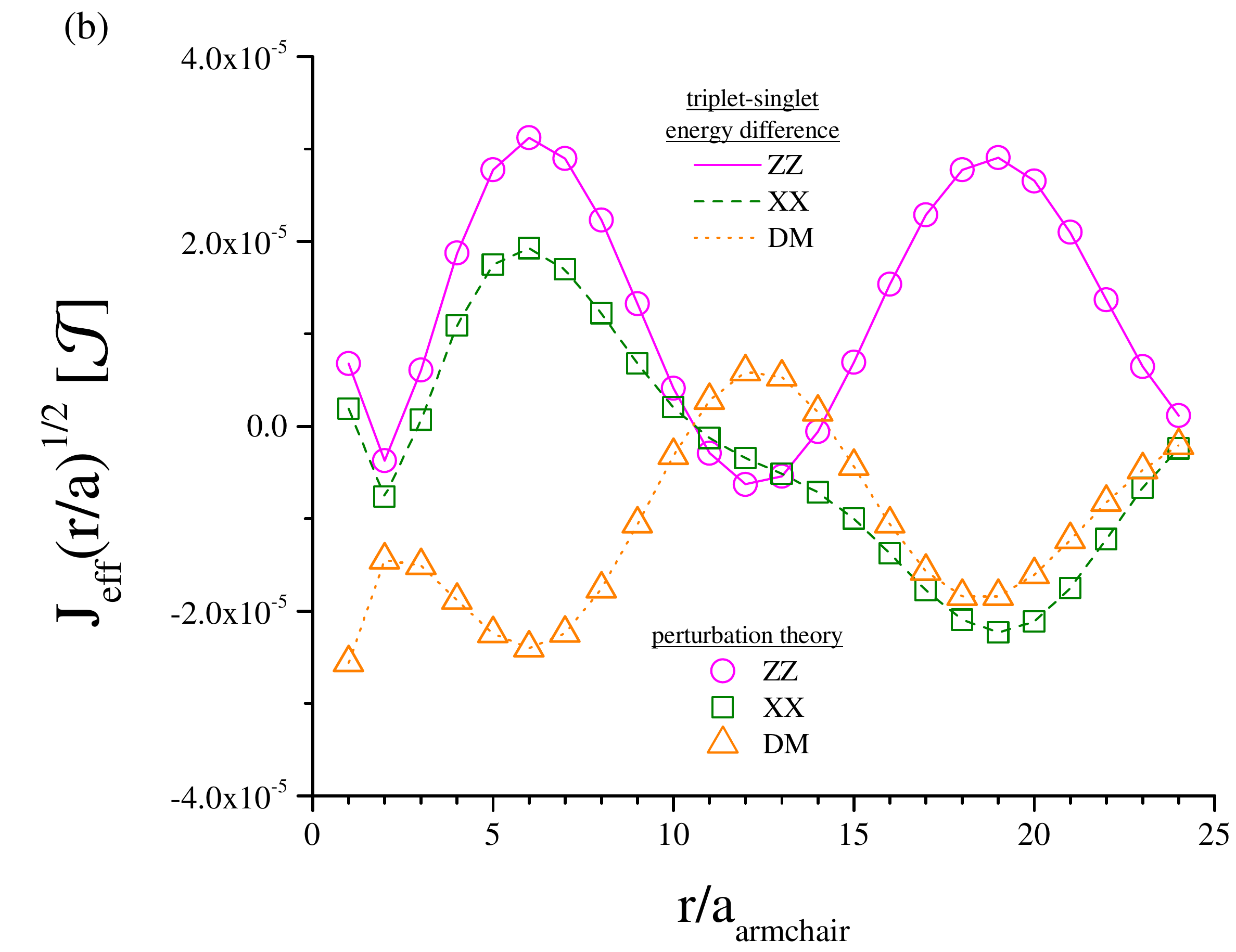}
\caption{RKKY interaction for impurities on a heteroribbon vs impurity separation, lying at the interface (a) zigzag on the Mo side, and (b) armchair on the W side. The interaction is calculated in units of ${\cal J} (=10$ meV), and scaled by the impurity separation $r^{1/2}$, normalized by the lattice constant along the chosen direction. Lines (empty symbols) indicate triplet-singlet energy difference (perturbation theory) results: magenta solid line (circle) for Ising $ZZ$, dark green dash line (square) for $XX$ and orange dot line (triangle) for $DM$ terms in Eq.\ \ref{jeffective1}. The Fermi energies are (a) $E_{F}=0.845$ eV for impurities hybridized to the zigzag interface, and (b) $E_{F}=0.799$ eV for the armchair interface.}
\label{Fig7}
\end{figure}

Let us now calculate the RKKY interaction between two magnetic impurities connected to a TMD HS interface, focusing on the exponent of the envelope and features of the FM/AFM oscillation. We consider a MoS$_{2}$-WS$_{2}$ heteroribbon with zigzag N100H30 ($\sim$30 nm interface) and armchair N30H100 ($\sim$28 nm interface) edges, hosting the two impurities in atomic lines at the interface between both materials. The Hamiltonian of the impurities, $H_{\text{I}}$, is added to the full MoS$_{2}$-WS$_{2}$ HS in Eq.\ \ref{heterolattice1}, so that $H = H_{\text{3OTB}} + H_{\text{I}}$. Here
\begin{equation}\label{impurities1}
  H_{\text{I}}=\mathcal{J} \sum_{i=1,2} \textbf{S}_{i}\cdot\textbf{s}_{\alpha_i}(\textbf{l}_i),
\end{equation}
with local exchange coupling ${\cal J}$ between the impurity spin $\textbf{S}_{i}$ and electrons in orbital $\alpha_i$ at the location of the impurity $\textbf{l}_{i}$. The electron spin density, at the sites where the impurities are hybridized, is
\begin{equation}\label{impurities2}
  \bm{s}_{\alpha}(\bm{l})=\frac{1}{2}\sum_{s,s'}d_{\alpha,s}^{\dagger}(\bm{l}) \bm{\sigma}_{s,s'}d_{\alpha,s'}(\bm{l}),
\end{equation}
where $\boldsymbol{\sigma}$ is the vector of spin-$\frac12$ Pauli matrices. After integration of the electronic degrees of freedom, one gets the inter-impurity effective exchange interaction as
\begin{eqnarray}\label{jeffective1}
H_{RKKY} &=& J_{XX}\left(S_{1}^{x}S_{2}^{x}+S_{1}^{y}S_{2}^{y}\right)+J_{ZZ}S_{1}^{z}S_{2}^{z}\nonumber\\
 & &+J_{DM}\left(\textbf{S}_{1}\times \textbf{S}_{2}\right)_{z},
\end{eqnarray}
where  $J_{XX} = J_{YY}$ (in-plane), $J_{DM}$ (in-plane Dzyaloshinskii-Moriya) and $J_{ZZ}$ (Ising) terms are proportional to the static spin susceptibility tensor of the electron system \cite{RudermanKittel1954,Kasuya1956,Yosida1957,Imamura2004}. Each of these $J$'s are effective coupling constants which will control the impurity interaction. We will jointly call them $J_{\mathrm{eff}}$ for simplicity. They are calculated in two ways: i) considering the energy difference between triplet and singlet impurity configurations after diagonalization of the full Hamiltonian $H$, and ii) second order perturbation theory. Both of these methods are explained in Appendix \ref{app:rkkycalculation}.

The bare couplings between localized and itinerant magnetic moments are set to ${\cal J}=10\text{ meV}$, in agreement with suggested exchange values between TMD and magnetic impurities \cite{ShiReview2018}. Additionally, we select midgap Fermi levels to reach states where the interfacial wave function is strong, such as $E_{F}=0.845$ eV for the zigzag [Fig.\ \ref{Fig2}(a)], and $E_{F}=0.799$ eV for the armchair interfacial states [Fig.\ \ref{Fig5}(a)]. Lastly, we assume hybridization to the $\alpha=d_{z^2}$ orbital of the metal Mo and W atoms, which has the largest amplitude for midgap levels, as shown in Fig.\ \ref{Fig2}(a) and \ref{Fig5}(a). Hybridization to other orbitals reduces the strength of the $J_{\mathrm{eff}}$ but it results in similar separation dependence.

Figure\ \ref{Fig7} shows representative RKKY interactions vs atomic separations $r/a$ between two magnetic impurities hybridized to sites at the interface and for midgap Fermi levels, for both zigzag [Fig.\ \ref{Fig7}(a)] and armchair [Fig.\ \ref{Fig7}(b)] interfaces. Results obtained numerically by the triplet-singlet energy difference approach [Appendix \ref{triplesinglet}] and analytically by second-order perturbation theory [Appendix \ref{pertub}] are seen to be in full agreement. The first impurity is placed at the interface at $r/a=0$ in these graphs, while the second is at $r/a\geq1$, up to $r/a_{\mathrm{zigzag}}=50$ and $r/a_{\mathrm{armchair}}=25$, which are the largest separations (half of the heteroribbon length) given the periodic boundary conditions; for the second half of each heteroribbon the RKKY is mirrored. Both panels show the two main features of the interaction explained before, the oscillatory form and the decaying envelope. The interaction oscillates between ferromagnetic ($J_{\mathrm{eff}}<0$) and antiferromagnetic ($J_{\mathrm{eff}}>0$) coupling between magnetic impurities, depending on their separation $r/a$. More importantly, the interaction is seen to decay as $J_{\mathrm{eff}}\propto1/r^d$. The effective dimensionality $d$ of the system, in both high symmetry directions, is found to be $d \simeq 1/2$, so that a long range interaction between impurities is effectively mediated by the HS interface. [Notice Fig.\ \ref{Fig7} shows $J_{\rm eff}\, r^{1/2}$ with no obvious remnant decaying behavior.]
Moreover, as described before in bulk 2D TMDs\cite{Parhizgar2013,Mastrogiuseppe2014}, and ribbon edges\cite{AvalosOvandoJPCM2018}, the strong spin-orbit interactions in the host result in sizable DM non-collinear interaction amplitudes between impurities.  These are comparable to the usual collinear interactions and result in interesting ground state configurations, as we will see.

Detailed inspection of the interaction curves for these Fermi levels indicates that the relative orientation of the second impurity moment changes with respect to the first as their separation increases. For the zigzag case, the $J_{XX}$ and $J_{DM}$ oscillation periods are nearly 3 sites and have a clear beating pattern, while for $J_{ZZ}$ the period is about 30 sites. Although $J_{XX}$ and $J_{DM}$ change sign every 3 sites, $J_{ZZ}$ is mostly AFM (positive) in nature, with a small FM window for $29\leq r/a_{\mathrm{zigzag}}\leq37$ and $r/a_{\mathrm{zigzag}}\leq8$. This behavior results in drastic variations in ground state spin orientations, depending on the separation $r/a_{\mathrm{zigzag}}$. For example, for $r/a_{\mathrm{zigzag}}=19$ all three terms have the same positive sign (AFM) and nearly the same amplitude, meaning that an isotropic Heisenberg-like interaction competes with a strong non-collinear DM term. On the other hand, at $r/a_{\mathrm{zigzag}}=21$ (AFM $J_{ZZ}$ and FM $J_{XX}$-$J_{DM}$) the low-energy configuration has both impurities pointing towards +\^{x} say, but one pointing towards ($-$\^{z}) and the other to (+\^{y},+\^{z}), a totally non-collinear arrangement. For the armchair case, all couplings have periods of about 15 sites. $J_{ZZ}$ is mostly AFM and $J_{DM}$ is mostly FM, except when $11<r/a_{\mathrm{armchair}}<14$, when $J_{ZZ}$ also changes to FM. This means that, for $r/a_{\mathrm{armchair}}\leq10$ one impurity may be pointing along ($-$\^{x},$-$\^{z}) and the other at (+\^{x},+\^{y},+\^{z}); for $11\leq r/a_{\mathrm{armchair}}\leq14$ the impurities would be along
($\pm$\^{x},+\^{z}), with one pointing towards $-$\^{y}; for $r/a_{\mathrm{armchair}}\geq15$, as $J_{XX}$ changes sign, both impurities point towards (+\^{x},$\pm$\^z). Other midgap interfacial Fermi levels we analyzed show similar general features with different magnitudes of the interaction.

We should comment on the role of the interface spatial structure in the interaction. Although we have chosen the impurities to lie on atomic positions where the orbital amplitude of the interfacial states is large, we see that nearby sites to the interface with smaller wave function amplitudes also result in similar RKKY interaction pattern, but with much smaller magnitude. In general, when one impurity lies on the strong wave function side (as shown in Fig.\ \ref{Fig7}) and the other on the other side of the interface, the interaction is decreased by half. When both sit away from the maximum wave function, the interaction is decreased by up to a factor of ten, but always exhibiting similar periodicity and oscillations. Analysis of different midgap doping levels leads to the same conclusions.

It is clear that the doping at midgap interfacial Fermi levels and suitable separation of impurities provides great tunable control on the resulting relative orientations of their magnetic moments. When impurities are set deep on the bulk monolayer, it has been shown that $d=2$\cite{Parhizgar2013,Mastrogiuseppe2014}, while for the HS interface here, we find indubitably $d\lesssim1$, highlighting the 1D character of the interfacial region. The fact that $d \lesssim 1$ results in effectively long-range interaction between impurities and it is likely related to the far-neighbor hopping described in Subsection \ref{subsec:ZigzagInterfaceStates}, as well as to the strong localization of the states at the interfacial region, as predicted in 1D quantum wires with Rashba SOC\cite{Silva2018unpublish}. The combination of state-of-the-art commensurate TMD interfaces\cite{Sahoo2018}, and magnetic impurities deposition and magnetic interaction measurements\cite{Khajetoorians2016,Steinbrecher2018}, could result in lateral TMD HS being promising tunable magnetic platforms to explore long-range magnetic interactions.

\section{Conclusions}
\label{sec:conclusions}
We have built theoretical models for studying pristine interfaces, both zigzag and armchair, between two different transition metal dichalcogenides, by using realistic tight-binding calculations and system sizes. We have shown that these interfaces can behave as 1D states with interesting features, including strong spin-orbit coupling. They can serve, for example, as unique effective hosts for the RKKY interaction between two magnetic impurities hybridized at the interface. Our numerical model is based on a successful three-orbital tight-binding model for describing pristine TMDs, which we modify to describe MoS$_{2}$-WS$_{2}$ and MoSe$_{2}$-WSe$_{2}$ planar heterostructures. We have characterized the dispersion curves from fully numerical diagonalization and proposed analytical expressions for these dispersions. We find that effective long-range hopping and spin-orbit interactions are necessary for a full description of these interface states. The RKKY interaction between two magnetic impurities at the interface results in long-range sub-1D interactions, showing that the interface states can behave as unusual 1D hosts and yield interesting physical behavior. This interfacial 1D platform could be used for the theoretical study of interface excitons\cite{Huang2014,Duan2014,Gong2014,Lau2018Arxiv}, p-n diodes and photodiodes\cite{Duan2014,Gong2014,Sahoo2018}, 1D quantum wells\cite{Huang2014} or charge density waves\cite{Giamarchi2004} in available heterostructures.

\begin{acknowledgments}
We acknowledge support from NSF grant DMR 1508325. O. \'A.-O. acknowledges a research fellowship from the Condensed Matter and Surface Science program at Ohio University.
\end{acknowledgments}

\bibliographystyle{apsrev4-1}
\bibliography{AvalosOvandoEtAl}

\newpage
\appendix

\section{RKKY effective interaction calculation}\label{app:rkkycalculation}

Here we present the two methods used for calculating the effective RKKY exchange interaction terms, $J_{XX}$, $J_{ZZ}$ and $J_{DM}=J_{XY}$ of Eq.\ \ref{jeffective1}, between two magnetic impurities in TMDs. In these calculations ${\cal J}$ is set as a constant.

\subsection{Triplet-Singlet Energy Difference}\label{triplesinglet}

The most accurate effective RKKY interaction is obtained from direct calculation of the difference between triplet and singlet impurity configurations in the system ground state as\cite{Black2010}
\begin{equation}\label{jeffective2}
  J_{\beta\beta'}^{\alpha_1,\alpha_2}= 2 \left[E(\uparrow_{\beta},\uparrow_{\beta'})-E(\uparrow_{\beta},\downarrow_{\beta'})\right],
\end{equation}
where $\alpha_1$ and $\alpha_2$ represent the orbitals to where the first and second impurities are hybridized, respectively. $\beta$ ($\beta'$) $\in\{X, Y, Z\}$ represents the direction of the spin projection for the first (second) magnetic impurity, for obtaining each of the $J$'s of Eq.\ \ref{jeffective1}.
The total system energies with magnetic impurities included in Eq.\ \ref{jeffective2}, are given by the sum of the sorted energy states of the full Hamiltonian up to a given Fermi energy $\epsilon_{\text{F}}$,
$E(s_{\beta},s_{\beta'})=\sum_{\mathrm{spin}=\uparrow,\downarrow}\sum_{i=1}^{\epsilon_{\text{F}}}\epsilon_{i,\mathrm{spin}}$,
as obtained after numerical diagonalization.

\subsection{Perturbation theory}\label{pertub}
The effective RKKY terms, $J$'s of Eq.\ \ref{jeffective1}, can also be calculated with second order perturbation theory \cite{Mattis, Nolting}, for small ${\cal J}$ in Eq.\ \eqref{impurities1}. Details of this method can be found in \onlinecite{AvalosOvando2016PRB}. We can rewrite Eq.\ \eqref{impurities1} as
\begin{equation}\label{impurities1_2}
H_{\text{I}}= \mathcal{J} \sum_{i=1,2}  S_{i}^z s^z_{\alpha_i}(\bm{l}_i) + \frac12\left[S_{i}^+ s^-_{\alpha_i}(\bm{l}_i) + S_{i}^- s^+_{\alpha_i}(\bm{l}_i)\right],
\end{equation}
with
\begin{align}\label{impurities2_2}
\begin{split}
s^z_{\alpha}(\bm{l}_j)&=\frac{1}{2}\left[d_{\alpha,\uparrow}^{\dagger}(\bm{l}_j) d_{\alpha,\uparrow}(\bm{l}_j) - d_{\alpha,\downarrow}^{\dagger}(\bm{l}_j) d_{\alpha,\downarrow}(\bm{l}_j)\right],\\
s^+_{\alpha}(\bm{l}_j)&=d_{\alpha,\uparrow}^{\dagger}(\bm{l}_j) d_{\alpha,\downarrow}(\bm{l}_j),\\
s^-_{\alpha}(\bm{l}_j)&=d_{\alpha,\downarrow}^{\dagger}(\bm{l}_j) d_{\alpha,\uparrow}(\bm{l}_j).
\end{split}
\end{align}

Then, by changing basis for the one that diagonalizes Eq.\ \ref{heterolattice1}, the spin operators are
\begin{align}\label{eq:spin_diag}
\begin{split}
s^z_{\alpha}(\bm{l}_j)&=\frac{1}{2}\sum_{\mu,\mu'}\left[\psi_{k,\mu}^* \psi_{k,\mu'} c_{\mu,\uparrow}^{\dagger} c_{\mu',\uparrow} - \psi_{k,\mu} \psi_{k,\mu'}^* c_{\mu,\downarrow}^{\dagger} c_{\mu',\downarrow}\right],\\
s^+_{\alpha}(\bm{l}_j)&=\sum_{\mu,\mu'} \psi_{k,\mu}^* \psi_{k,\mu'}^* c_{\mu,\uparrow}^{\dagger} c_{\mu',\downarrow},\\
s^-_{\alpha}(\bm{l}_j)&=\sum_{\mu,\mu'} \psi_{k,\mu} \psi_{k,\mu'} c_{\mu,\downarrow}^{\dagger} c_{\mu',\uparrow},
\end{split}
\end{align}
where $\psi_{k,\mu}$ is the component of the eigenvector for site $j$, orbital $\alpha$, and spin projection $s$; $c(c_{\mu,s}^{\dagger})$ are the anihilation (creation) operators in the diagonal basis.\cite{AvalosOvando2016PRB} The  second order correction to the energy in perturbation theory is given by
\begin{equation}\label{eq:perturb}
E^{(2)} = \sum_{ex,\mathcal{D}'}\frac{|\Braket{GS;\mathcal{D}|H_I|ex;\mathcal{D}'}|^2} {E_{GS} - E_{ex}}.
\end{equation}
In this expression, $\ket{GS;\mathcal{D}}\equiv \ket{GS}\ket{\mathcal{D}}$, where $\ket{GS}$ is the ground state of the new basis Hamiltonian and $\ket{\mathcal{D}}$ the ground state spin configuration of the two disconnected magnetic moments. Similarly, $\ket{ex}$ denote particle-hole excitations of the electron gas, and $\ket{\mathcal{D}'}$ are excited configurations of the two impurities.
Using \eqref{eq:spin_diag} and \eqref{impurities1_2}, one can write \eqref{eq:perturb} as
\begin{equation}
\begin{split}
E^{(2)}_{\alpha,\alpha'} = \frac{\mathcal{J}^2}{2}&\sum_{\substack{\mu \leq \mu_F\\\mu'>\mu_F}}\frac{1}{\epsilon_\mu -\epsilon_{\mu'}}\bra{\mathcal{D}}
J_{ZZ}^{\alpha,\alpha'}(\bm{r}_j,\mathbf{r}_{j'})  S_j^z S_{j'}^z \\
&+ J_{XX}^{\alpha,\alpha'}(\bm{r}_j,\mathbf{r}_{j'})  (S_j^x S_{j'}^x + S_j^y S_{j'}^y) \\
&+ J_{XY}^{\alpha,\alpha'}(\bm{r}_j,\mathbf{r}_{j'})  (S_j^x S_{j'}^y - S_j^y S_{j'}^x)\ket{\mathcal{D}}
\end{split}
\end{equation}
with the effective $J$'s of Eq.\ \ref{jeffective1} given by
\begin{align}
\begin{split}
J_{ZZ}^{\alpha,\alpha'} (\bm{l}_j,\mathbf{l}_{j'})& = \sum_{\substack{\mu \leq \mu_F\\\mu'>\mu_F}}\text{Re} \left[ (\psi_{j, \mu}^\alpha)^* \psi_{j, \mu'}^\alpha \psi_{j', \mu}^{\alpha'} (\psi_{j', \mu'}^{\alpha'})^*\right],\\
J_{XX}^{\alpha,\alpha'}(\bm{l}_j,\mathbf{l}_{j'})& = \sum_{\substack{\mu \leq \mu_F\\\mu'>\mu_F}}\text{Re} \left[ \psi_{j, \mu}^\alpha \psi_{j, \mu'}^\alpha (\psi_{j', \mu}^{\alpha'})^* (\psi_{j', \mu'}^{\alpha'})^*\right],\\
J_{XY}^{\alpha,\alpha'}(\bm{l}_j,\mathbf{l}_{j'})& = -\sum_{\substack{\mu \leq \mu_F\\\mu'>\mu_F}}\text{Im} \left[ \psi_{j, \mu}^\alpha \psi_{j, \mu'}^\alpha (\psi_{j', \mu}^{\alpha'})^* (\psi_{j', \mu'}^{\alpha'})^*\right].
\end{split}
\end{align}
$\mu_F$ denotes the level index associated with a given Fermi energy $\varepsilon_{F}$ (in eV) in the TMD ribbon, and correspond to the amount of \emph{p}-doping or gating within the band gap of the 2D bulk.

\end{document}